\documentclass[pdflatex,sn-mathphys-num]{sn-jnl}

\usepackage{graphicx}%
\usepackage{multirow}%
\usepackage{amsmath,amssymb,amsfonts}%
\usepackage{amsthm}%
\usepackage{mathrsfs}%
\usepackage[title]{appendix}%
\usepackage{hyperref}
\usepackage{xcolor}%
\usepackage{subcaption}
\usepackage{textcomp}%
\usepackage{manyfoot}%
\usepackage{booktabs}%
\usepackage{algorithm}%
\usepackage{algorithmicx}%
\usepackage{algpseudocode}%
\usepackage{listings}%
\usepackage{hyperref}
\usepackage{algorithm}
\usepackage[normalem]{ulem}
\DeclareMathOperator*{\argmin}{arg\,min}
\usepackage{siunitx}
\usepackage{multirow}
\usepackage{gensymb}
\usepackage{setspace}

\theoremstyle{thmstyleone}%
\theoremstyle{thmstyletwo}%

\theoremstyle{thmstylethree}%

\begin{document}

\title[Article Title]{Cluster-Aware Conformal Calibration for Spatio-Temporal Distributional Prediction}


\author[1]{\fnm{Gooyoung} \sur{Kim}} 
\email{kgy9023@snu.ac.kr}

\author[1]{\fnm{Chae Young} \sur{Lim}}
\email{twinwood@snu.ac.kr}

\author[2]{\fnm{Wen-Ting} \sur{Wang}}
\email{egpivo@gmail.com}

\author*[3]{\fnm{Hao-Yun} \sur{Huang}}\email{hhuscout@gms.ndhu.edu.tw}

\author*[3]{\fnm{Wei-Ying} \sur{Wu}}\email{wuweiying1011@gmail.com}

\affil[1]{\orgdiv{Department of Statistics}, \orgname{Seoul National University}, \orgaddress{\city{Seoul}, \postcode{08826}, \country{South Korea}}}

\affil[2]{\orgdiv{Institution of Statistics}, \orgname{National Chung Hsing University}, \orgaddress{\city{Taichung City}, \postcode{402}, \country{Taiwan}}}

\affil[3]{\orgdiv{Department of Applied Mathematics}, \orgname{National Dong Hwa University}, \orgaddress{\city{Hualien}, \postcode{974},  \country{Taiwan}}}


\abstract{DeepKriging-style models, such as Spatio-Temporal DeepKriging, improve scalability through basis-function embeddings and stochastic gradient learning; however, fixed regular-grid spatial bases remain inefficient under highly non-uniform sampling patterns, often over-allocating capacity to sparse regions while under-resolving dense clusters. To address this limitation, we propose a practical extension of DeepKriging for reliable spatio-temporal distributional forecasting, incorporating cluster-adaptive spatial bases - whose centers and scales are initialized from {the spatial sampling density} - to better capture heterogeneous spatial sampling, together with cluster-aware conformal calibration that determines prediction-interval widths within spatial clusters (with a global fallback when calibration samples are insufficient). The resulting calibration pipeline explicitly targets spatial heterogeneity and local miscalibration, and experiments, including simulation studies and PM$_{2.5}$ data analysis, demonstrate substantially improved coverage accuracy and tail reliability under clustered observation patterns compared with a global conformal baseline.}

\keywords{Deep learning,  spatio-temporal, kriging, cluster, conformal}



\maketitle

\section{Introduction}\label{sec1}

Spatio-temporal interpolation plays a fundamental role in environmental monitoring, climate modeling, and sensor networks, where observations are often irregularly sampled. Classical kriging provides strong theoretical guarantees, including the best linear unbiased predictor (BLUP) and predictive distributions under Gaussian assumptions. However, classical kriging methods often suffer from substantial computational burdens for large spatial or spatio-temporal datasets and typically rely on restrictive covariance assumptions \cite{Sun2012}. Comprehensive reviews of space--time covariance structures and covariance-based modeling approaches are provided in \cite{Montero2015, Chen2021}.

Beyond covariance-based approaches, numerous methods have been proposed for modeling stationary and nonstationary spatio-temporal processes, including hierarchical Bayesian frameworks, mixture-based covariance constructions, deformation methods, and dynamic spatio-temporal models \cite{Wikle1998, Stroud2001, Ma2002, Huang2004, Kolovos2004, Stein2005, Fuentes2008, Bruno2009, Bartlett2013, Sigrist2012, Xu2018}. A broader overview of statistical methodologies for stationary and nonstationary spatio-temporal data can be found in \cite{Cressie2015}.

In recent years, deep neural networks (DNNs) have been increasingly adopted in environmental sciences and spatio-temporal analysis.  A comprehensive review of recent statistical and deep learning frameworks for spatio-temporal forecasting is provided by \cite{Wikle2023}.  Specifically, DeepKriging (DK) embeds spatial coordinates
via basis expansion and feeds the resulting vectors, together with covariates,
into a deep neural network (DNN) to model the spatial field nonlinearly while reflecting spatial dependence ~\cite{chen2024deepkriging, LinHuangTzeng2023,nag2023stdk, Nag2025}.

DeepKriging-style models, including Spatio-Temporal DeepKriging (STDK), improve scalability by embedding spatial and temporal coordinates through basis functions and learning nonlinear predictors via stochastic gradient descent~\cite{nag2023stdk}, but fixed regular-grid spatial bases are inefficient under highly non-uniform sampling: they over-allocate capacity to sparse areas while under-resolving dense clusters.
These challenges become more pronounced under nonstationarity, clustered sampling, and time-varying observation masks, where both local variability and uncertainty can change sharply over space and time. In classical spatial and spatio-temporal geostatistics, prediction intervals are derived from model-based predictive distributions under assumed covariance structure~\cite{cressie1993spatial, cressie2011spatiotemporal}. Although principled, such intervals can be miscalibrated when covariance assumptions are misspecified or heterogeneity is strong.

In this paper, we address this mismatch with \textbf{cluster-adaptive spatial bases} and emphasize \textbf{distributional forecasting} and \textbf{reliable uncertainty quantification}. Our contributions are:
\begin{itemize}
    \item A cluster-adaptive spatial basis with learnable centers and scales, initialized by the spatial sampling density and optimized jointly with network weights.
    \item A multi-quantile joint training extension with non-crossing regularization for distributional forecasting.
    \item A cluster-aware conformal calibration layer that adjusts interval widths per spatial cluster (global fallback for small clusters), improving coverage under heterogeneous sampling.
\end{itemize}
Conceptually, our design follows a density-estimation intuition: the spatial sampling density guides spatial capacity allocation before task-specific supervised refinement. Unlike standalone density estimation, our target is calibrated conditional quantile prediction over both space and time.
Our implementation is publicly available at \url{https://github.com/STLABTW/da-stdk}.

\section{Methodology}\label{Methodology}
Let $\{Y(s,t): s\in\mathcal{S}\subset\mathbb{R}^2,\ t\in\mathcal{T}\}$ denote a spatio-temporal process, and let $x(s,t)$ denote observed covariates. Our goal is to estimate conditional quantiles of $Y(s,t)$ under heterogeneous, potentially nonstationary, and spatially clustered observation patterns, while preserving temporal dynamics through explicit time-basis modeling.

At a high level, our method separates \emph{distribution-shape learning} and \emph{coverage calibration}: Sections~\ref{sec:method:basis}--\ref{sec:method:mq} define the {DA-STDK-MQ backbone} (adaptive spatial basis + multi-quantile learning), while Section~\ref{sec:method:conformal} applies conformal calibration to enforce interval reliability. The backbone keeps the same feed-forward design spirit as STDK (basis embeddings + covariates $\rightarrow$ neural network)~\cite{nag2023stdk}, while replacing fixed spatial bases with cluster-adaptive ones and a quadratic loss with a non-crossing multi-quantile loss. 
Compared with STDK, another key distinction lies in the full uncertainty pipeline, followed by a cluster-aware conformal adjustment to correct local coverage.

From a statistical perspective, this can be viewed as a density-informed representation step followed by supervised quantile fitting and post-hoc coverage correction.
{
Following the STDK input design, our predictor takes three feature groups as input: spatial bases $\phi(s;\eta)$, temporal bases $\psi(t)$, and observed covariates $x(s,t)$. These are concatenated and passed into a shared trunk network (Fig.~\ref{fig:backbone-flow}),
then mapped to multiple quantile outputs by quantile-specific heads (Section~\ref{sec:method:mq}).
\begin{figure}[t]
\centering
\includegraphics[width=0.9\textwidth]{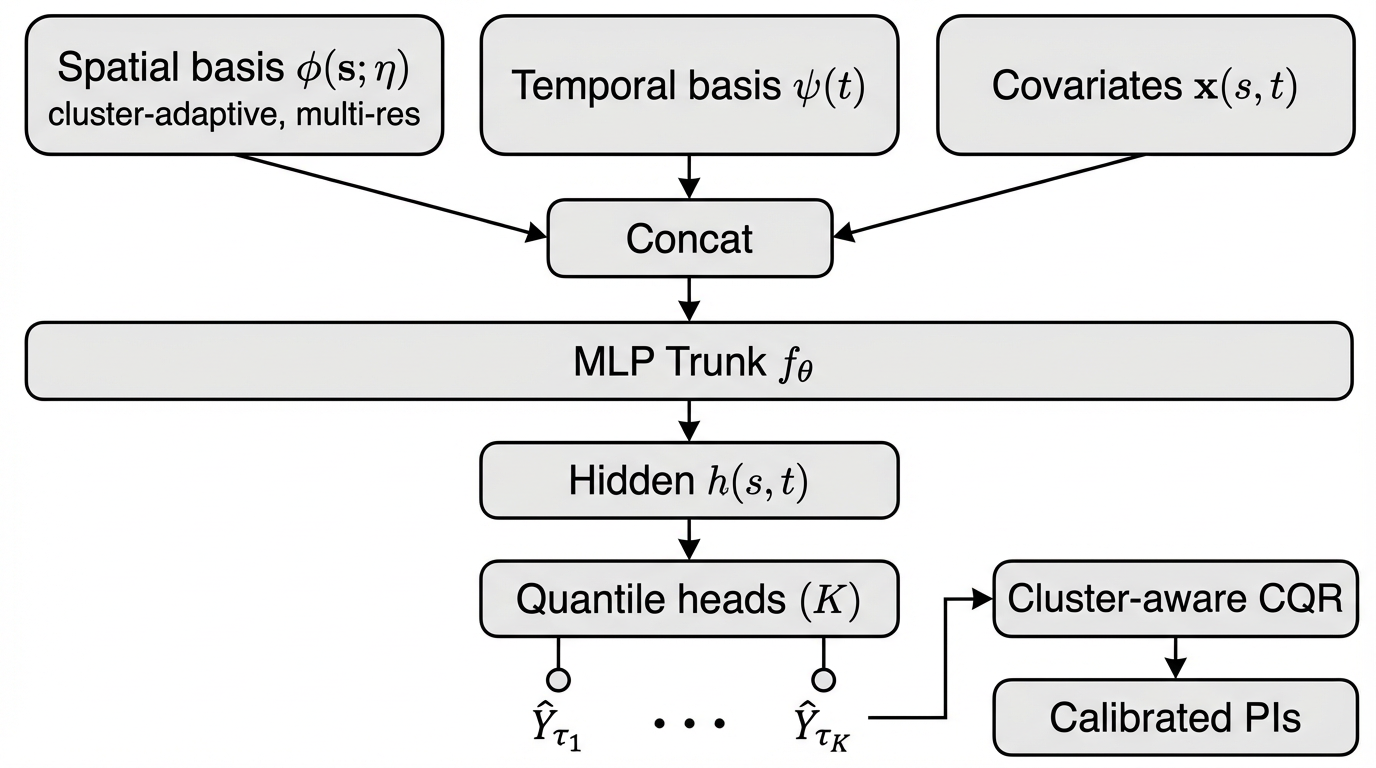}
\caption{DA-STDK-MQ backbone: concatenated input $[\phi(s;\eta),\psi(t),x(s,t)]$ $\to$ MLP trunk $f_\theta$ $\to$ hidden state $h(s,t)$ $\to$ quantile heads $\to$ $\{\hat{Y}_{\tau_k}\}_{k=1}^K$.}
\label{fig:backbone-flow}
\end{figure}
}

\subsection{Cluster-Adaptive Spatial Basis}\label{sec:method:basis}
We assume the spatial field contains multiple local clusters with different local sampling-density patterns and spatial heterogeneity. Our goal in this module is to \emph{learn where the representative centers should be} and then learn cluster-level spatial representations around them.

Before end-to-end training, we provide a weak, data-informed initialization for center locations so optimization starts from a plausible geometry rather than from arbitrary points. Let $s \in \mathcal{S} \subset \mathbb{R}^2$ be spatial coordinates and $t \in \mathcal{T}$ be time. We initialize centers via a density-weighted clustering objective:

\begin{equation}\label{eq:u-init}
\mathcal{U}_{\mathrm{init}}=\arg\min_{\{u_c\}_{c=1}^C}\sum_{i=1}^N w_i\min_{1\le c\le C}\|s_i-u_c\|_2^2.
\end{equation}
Here $C$ denotes the number of initialization centers. In our design, it follows the multi-resolution basis budget, i.e., $C=\sum_{\ell=1}^L K_\ell$, {where $K_\ell$ is the number of bases in the $\ell$ level}. In principle, $C$ can also be treated as a tunable hyperparameter; however, because this stage provides initialization only, we do not emphasize tuning $C$ in this study. Also, $w_i$ reflects local sampling density (larger in denser regions). We optimize Eq.~\eqref{eq:u-init} with balanced k-means; a { Gaussian mixture model (GMM)} initializer (component means) is a practical fallback when imbalance is severe, but it is outside the scope of this study.

The statistical rationale is that this initialization acts as a low-variance anchor for spatial partitioning under irregular sampling. In finite samples, fully free center learning from random starts can be unstable and may overfit sparse regions. Starting from density-informed centers provides a consistent local reference geometry, while subsequent gradient-based updates reduce initialization bias and adapt centers to the task loss.

Given (initialized and then trainable) centers, we build a \textbf{multi-resolution spatial basis} to learn cluster representations around each center:
\begin{equation}\label{eq:spatial-basis}
    {\phi(s;\eta) = \big[\phi^{(1)}(s;\eta^{(1)})^\top,\ldots,\phi^{(L)}(s;\eta^{(L)})^\top\big]^\top, \quad \phi^{(\ell)}(s;\eta^{(\ell)}) = \big( b(\|s-u_j^{(\ell)}\|/r_j^{(\ell)}) \big)_{j=1}^{K_\ell}},
\end{equation}
where $b(\cdot)$ is a radial kernel (e.g., Wendland), and {$ \eta=(\eta^{(\ell)})^L_{\ell=1}$, where $\eta^{(\ell)}=\{(u_j^{(\ell)},r_j^{(\ell)})\}_{j=1}^{K_\ell}$ denotes level-$\ell$ centers and scales. The basis counts $\{K_\ell\}_{\ell=1}^L$ are \emph{predefined} hyperparameters (fixed before training), so centers are allocated level-by-level.

Consider the deep neural network with the input layer $\big[x(s,t),\phi(s;\eta),\psi(t)\big]$, where 
$x(s,t)$ denotes observed covariates, $\phi$'s are the multi-resolution spatial basis with the parameter $\eta$ defined in (\ref{eq:spatial-basis}), and $\psi(t)$'s are temporal basis functions such as Gaussian
radial basis functions. The corresponding shared trunk output is $h(s,t;\theta,\eta) \in \mathbb{R}^d$  with the trunk parameters ($\theta$) and basis parameter ($\eta$). }
We set initial scales from local neighbor distances so each center has a locally adaptive bandwidth. Coarser levels capture global structure, while finer levels capture local variation. Accordingly, the spatial-basis initialization $\eta^{\mathrm{init}}$ is built from $\mathcal{U}_{\mathrm{init}}$ and local-distance scales. The corresponding trained center set is
\[
\mathcal{U}_{\mathrm{tr}}=\{u_c\}_{c=1}^C.
\]

During training, all basis parameters $\eta^{(\ell)}$ are optimized jointly with network weights, but updates are constrained to stay close to the initialized center geometry. For center stability, one option is an explicit proximal penalty (e.g., $\|u-u^{\text{init}}\|^2$)~\cite{parikh2014proximal,li2020federated}. Instead, we use \emph{distance-dependent} gradient damping: centers that drift farther from initialization receive smaller gradients. This keeps adjustment conservative and improves optimization stability; damping is controlled by $d_{\text{th}}$ and $\kappa$ (Algorithm~\ref{alg:overall-opt}, step (b.3)). To keep centers inside the spatial domain, we use
\begin{equation}
    P_{\text{domain}} = \sum_{\ell=1}^L \sum_{j=1}^{K_\ell} \sum_{d=1}^{2}
    \left[\max(0,-u_{j,d}^{(\ell)})^2 + \max(0,u_{j,d}^{(\ell)}-1)^2\right],
\end{equation}
which penalizes centers outside $[0,1]^2$. This term is included in training with coefficient $\lambda_{\text{domain}}$.

\subsection{Distributional Forecasting via Multi-Quantile Joint Training}\label{sec:method:mq}

Let $y=Y(s,t)$ denote the response at location $(s,t)$. 
For given target quantile levels $0<\tau_1<\cdots<\tau_K<1$, the $\tau_k$-th quantile predictor is defined as
\begin{equation}\label{eq:mq-head}
\hat{Y}_{\tau_k}(s,t) = [1, h(s,t;\theta,\eta)^\top] \beta_k.
\end{equation}

Inspired by \cite{moon2021learning}, to prevent the quantile crossing issue, we adopt the reparameterization 
\[
\beta_k = \sum_{\ell=1}^k \delta_\ell,
\]
where $\delta_1 = \beta_1$ and $\delta_k = \beta_k - \beta_{k-1}$ for $k \geq 2$. 

Let $P_{nc}(\delta)$ denote the parameter-level non-crossing regularizer that enforces the feasibility condition, defined as
\begin{equation}
P_{nc}(\delta)
=
\sum_{k=2}^K J(\delta_k),
\qquad
J(\delta_k)
=
\delta_{k,0}
-
\max\bigg(
\delta_{k,0},
\sum_{j=1}^d \max(0,-\delta_{k,j})
\bigg),
\end{equation}
where $\delta_{k,0}$ corresponds to the intercept component and $\delta_{k,j}$ ($j=1,\ldots,d$) denotes the coefficient associated with the $j$-th feature.

The multiple quantiles $\{\hat{Y}_{\tau_k}(s,t)\}_{k=1}^K$ are jointly estimated by solving
\begin{equation}\label{eq:overall-objective}
\argmin_{\theta,\eta,\delta} \mathcal{L}_{\text{overall}}
:=
\sum_{k=1}^K \rho_{\tau_k}\big(y-\hat{Y}_{\tau_k}(s,t)\big)
+ \lambda_{\text{domain}}\,P_{\text{domain}}
+ \lambda_{\text{nc}}\,P_{nc}(\delta),
\end{equation}
where $\rho_{\tau_k}(\cdot)$ denotes the quantile loss at level $\tau_k$.

We initialize $\eta$ using the density-aware centers and local distance scales described in Section~\ref{sec:method:basis}. 
The model parameters  including the trunk network parameters, the quantile-head increments (with violation-set-aware non-crossing adjustments), and the spatial centers are iteratively updated. Distance-dependent gradient damping is applied to stabilize the optimization process.
The detailed training updates are summarized in Algorithm~\ref{alg:overall-opt}.

\begin{algorithm}[H]
\caption{Overall optimization of DA-STDK-MQ }
\label{alg:overall-opt}
\begin{enumerate}
    \item[(a)] Initialize $\eta\leftarrow\eta^{\mathrm{init}}$ from $\mathcal{U}_{\mathrm{init}}$ (Eq.~\eqref{eq:u-init}, density-aware centers) and local-distance scales.
    \item[(b)] For $t=1,\dots,T$, do:
    \begin{enumerate}
        \item[(b.1)] Update trunk parameters:
        \[
        \theta \leftarrow \theta-\gamma_t\nabla_\theta\mathcal{L}_{\text{overall}}.
        \]
        \item[(b.2)] \textbf{L1-penalization non-crossing update} (Moon et al., 2021):
        \begin{enumerate}
            \item[(b.2.1)] Define the violation set:
            \[
            A(\delta)\leftarrow\left\{k\ge2:\delta_{k,0}<\sum_{j=1}^{d}\max(0,-\delta_{k,j})\right\}.
            \]
            \item[(b.2.2)] Update quantile-head increments:
            \[
            \delta_k \leftarrow \delta_k-\gamma_t
            \begin{cases}
            \nabla_{\delta_k}\mathcal{L}_{\text{overall}}+\lambda_{\text{nc}}\nabla_{\delta_k}J(\delta_k), & k\ge2,\ k\in A(\delta),\\[3pt]
            \nabla_{\delta_k}\mathcal{L}_{\text{overall}}, & \text{otherwise},
            \end{cases}
            \quad k=1,\dots,K.
            \]
        \end{enumerate}
        \item[(b.3)] \textbf{Distance-dependent gradient damping for centers} 
        
        {For each $\ell=1,\ldots,L$, $j=1,\ldots,K_\ell$}:
        \begin{enumerate}
            \item[(b.3.1)] Compute displacement and damping factor:
            \[
            d_j^{(\ell)}=\|u_j^{(\ell)}-u_{j,\mathrm{init}}^{(\ell)}\|_2,\quad
            a_j^{(\ell)}=\exp\!\big(-\kappa\max(0,d_j^{(\ell)}-d_{\text{th}})\big).
            \]
            {($j$ indexes the center within level $\ell$; $d_j^{(\ell)}$ is displacement, $a_j^{(\ell)}$ is damping factor.)}

            \item[(b.3.2)] Apply damped gradient and update center:
            \[
            \tilde{\nabla}_{u_j^{(\ell)}}\mathcal{L}_{\text{overall}}=a_j^{(\ell)}\nabla_{u_j^{(\ell)}}\mathcal{L}_{\text{overall}},\quad
            u_j^{(\ell)}\leftarrow u_j^{(\ell)}-\eta_u\tilde{\nabla}_{u_j^{(\ell)}}\mathcal{L}_{\text{overall}}.
            \]
        \end{enumerate}
    \end{enumerate}
\end{enumerate}
\end{algorithm}

\subsection{Cluster-Aware Conformal Calibration for Reliable Coverage}\label{sec:method:conformal}
While multi-quantile regression provides distributional estimates, it does not guarantee finite-sample coverage under spatial heterogeneity. We therefore apply conformalized quantile regression (CQR) as a post-processing step using a calibration split (validation when available; in our train\_ratio $=0.8$ setting, calibration uses the validation split; otherwise a held-out subset of training data).
Using the quantile outputs in Section~\ref{sec:method:mq} (Eq.~\eqref{eq:mq-head}), we first form nominal prediction intervals, then calibrate them by conformal adjustment.

\paragraph{Global CQR baseline}
As a baseline, we use standard conformalized quantile regression (CQR) with a single global adjustment term~\cite{romano2019cqr}. Given calibration data $\{(s_i,t_i,Y_i)\}_{i=1}^n$ and nominal miscoverage $\alpha$ (e.g., 0.1 for 90\% intervals), define nonconformity scores
\begin{equation}
    R_i = \max\big(\hat{Y}_{\alpha/2}(s_i,t_i)-Y_i,\;Y_i-\hat{Y}_{1-\alpha/2}(s_i,t_i),\;0\big).
\end{equation}
The global conformal adjustment is
\begin{equation}
    \hat{q}_{\text{global}} = Q_{1-\alpha}(\{R_i\}_{i=1}^n),
\end{equation}
which yields the global interval
\begin{equation}\label{eq:global-conformal}
    \hat{C}^{\text{global}}_{1-\alpha}(s,t)=\big[\hat{Y}_{\alpha/2}(s,t)-\hat{q}_{\text{global}},\;\hat{Y}_{1-\alpha/2}(s,t)+\hat{q}_{\text{global}}\big].
\end{equation}
This baseline is simple, but a single $\hat{q}_{\text{global}}$ can under-adjust hard regions and over-adjust easy regions when spatial uncertainty is heterogeneous.

\paragraph{Cluster-aware CQR}
To adapt calibration locally, we use a center set for calibration, denoted by $\mathcal{U}_{\mathrm{cal}}=\{u_c\}_{c=1}^C$. {For \textbf{DA-STDK-MQ }, the clusters are defined by the same spatial centers as in the cluster-adaptive basis (Section~\ref{sec:method:basis}): we set $\mathcal{U}_{\mathrm{cal}}=\mathcal{U}_{\mathrm{tr}}$ (trained centers) so that the partition of space is consistent with the backbone representation. }We then define nearest-center cluster assignment. This design is closely related to localized/group-conditional conformal ideas in the conformal prediction literature~\cite{guan2023localized}. Define
\begin{equation}\label{eq:cluster-assign}
    c(s)=\arg\min_{c}\|s-u_c\|
\end{equation}
and  
\begin{equation}\label{eq:cluster-conformal}
    \hat{C}^{\text{cluster}}_{1-\alpha}(s,t)=\big[\hat{Y}_{\alpha/2}(s,t)-\hat{q}_{c^\star},\;\hat{Y}_{1-\alpha/2}(s,t)+\hat{q}_{c^\star}\big], 
\end{equation}
where $c^\star=c(s)$ is the cluster assigned to the test point $(s,t)$ via Eq.~\eqref{eq:cluster-assign}, and $\hat{q}_{c^\star}$ is the conformal half-width for that cluster: the $(1-\alpha)$-quantile of nonconformity scores among calibration points in cluster $c^\star$, or $\hat{q}_{\mathrm{global}}$ if the cluster has too few points (Algorithm~\ref{alg:cluster-cqr}, step (d)).

\begin{algorithm}[t]
\caption{Cluster-aware conformal calibration}
\label{alg:cluster-cqr}
\begin{enumerate}
    \item[(a)] Assign calibration points by nearest trained center set $\mathcal{U}_{\mathrm{cal}}$ (Section~\ref{sec:method:basis}): $c_i=c(s_i)$.
    {($i$ indexes calibration points $1,\ldots,n$; $c_i$ is the cluster index for point $i$ via Eq.~\eqref{eq:cluster-assign}.)}
    \item[(b)] Compute
    \[
    R_i\leftarrow\max\!\big(\hat{Y}_{\alpha/2}(s_i,t_i)-Y_i,\;Y_i-\hat{Y}_{1-\alpha/2}(s_i,t_i),\;0\big).
    \]
    {($R_i$: nonconformity score for calibration point $i$.)}
    \item[(c)] Compute the global quantile:
    \[
    \hat{q}_{\mathrm{global}}\leftarrow Q_{1-\alpha}(\{R_i\}_{i=1}^n).
    \]
    {($n$: number of calibration points; $Q_{1-\alpha}$: $(1-\alpha)$-quantile.)}
    \item[(d)] For each cluster $c$, define
    \[
    \hat{q}_c=
    \begin{cases}
    Q_{1-\alpha}(\{R_i:c_i=c\}), & n_c\ge n_{\min},\\[2pt]
    \hat{q}_{\mathrm{global}}, & n_c<n_{\min}.
    \end{cases}
    \]
    {($c$ indexes clusters $1,\ldots,C$; $n_c=|\{i:c_i=c\}|$; $n_{\min}$: minimum sample threshold for per-cluster quantile.)}
    \item[(e)] For test point $(s,t)$ with $c^\star=c(s)$, output Eq.~\eqref{eq:cluster-conformal}.
    {($c^\star$: cluster assigned to the test point via nearest center in Eq.~\eqref{eq:cluster-assign}.)}
\end{enumerate}
\end{algorithm}

\section{Numerical Study}
\subsection{Experimental Setup}
This subsection reports model/training choices and the evaluation workflow used for both simulation and real-data studies.

\paragraph{Model and Training Settings.}
We use a 3-layer MLP with hidden sizes 256, 256, 128, ReLU activations, layer normalization, and dropout 0.1. Optimization uses AdamW with base learning rate 0.01, weight decay $5\times 10^{-4}$, and a reduced rate for basis parameters (0.0005), batch size 4096, maximum 500 epochs, and early stopping patience 50. Spatial bases use Wendland kernels with multi-resolution counts 25, 81, 121 (total 227). Temporal bases use Gaussian RBFs with 10, 15, 45 bases (total 70); temporal bandwidth follows the implementation default (2.5$\times$ grid spacing) and is fixed across methods. Spatial basis initialization uses balanced k-means for DA-STDK-MQ  and fixed grids for STDK. We use gradient damping for basis-center updates. For non-crossing, we use the $\ell_1$-penalization algorithm (penalty gradient applied only to heads in the violation set $A(\delta)$), and tune $\lambda_{\text{nc}}$ by grid search (candidate values include $0$, $10^{-3}\!\sim\!10^{-1}$ log-scale, and additional larger values). In our current KAUST runs, the best CRPS is achieved at $\lambda_{\text{nc}}=0$, so the reported results (Tables~\ref{tab:crps-latest} and~\ref{tab:coverage-latest}) use this selected value (i.e., no active non-crossing penalty term). For fair numerical comparison, the domain penalty is applied only to DA-STDK-{MQ}  (learnable centers); STDK uses fixed spatial bases and thus has no $P_{\text{domain}}$ term. In our latest run, DA-STDK-{MQ}  conformal calibration uses trained centers for cluster assignment, i.e., $\mathcal{U}_{\mathrm{cal}}=\mathcal{U}_{\mathrm{tr}}$. Overall, we isolate gains from spatial adaptation while keeping temporal modeling identical across methods.

\paragraph{Experimental Pipeline.}
We adopt a reproducible pipeline with fixed configurations and seeds: (1) data preparation with fixed dataset split and observation regime; (2) baseline STDK training; (3) DA-STDK-MQ  training with adaptive bases; (4) cluster-aware conformal calibration using a held-out subset (Ours); (5) evaluation of CRPS and coverage metrics with spatial visualizations.



\subsection{Simulation Data: KAUST Benchmark}
\label{sec:method:Simulation}
\paragraph{Data Description and Scenarios.}

We evaluate the performance of the proposed approach using the KAUST spatio-temporal competition datasets, which are generated from a zero-mean Gaussian process with a Matérn space–time covariance structure. The competition comprises six datasets (2a-7, 2a-8, 2a-9, 2b-7, 2b-8, 2b-9); further details can be found in \cite{Abdulah2022}.
For brevity, we report results only for the dataset 2b-8, as the findings for the remaining datasets are similar. Additional results are provided in the Appendix~\ref{secA1}.
The spatial domain consists of 10,000 locations for the dataset 2b-8, each observed over 100 time points.

In the simulation studies, we consider four observation regimes defined by crossing two factors: fixed vs. random observation sets over time, and uniform vs. clustered spatial sampling:

{{
\begin{enumerate}
  \item \textbf{Observation method:}
  \begin{itemize}
    \item \emph{Fixed}: The same set of spatial observation locations ($s$)  is used at each time step, simulating a permanent monitoring network.
    \item \emph{Random}: Observation sites are randomly resampled at each time step, with each $(t,s)$ observed independently via Bernoulli sampling; consequently, the set of observed sites varies over time.
  \end{itemize}
  \item \textbf{Spatial sampling pattern:}
  \begin{itemize}
    \item \emph{Uniform}: Spatial observation locations are sampled uniformly at random over the spatial domain.
    \item \emph{Clustered}: Spatial observation locations are sampled with a bias toward one corner of the domain (via a Gaussian kernel), inducing spatial imbalance. In our simulation studies, under the clustered setting, locations are sampled from a density proportional to $(1 + 10\|s\|)^{-2}$ on $[0,1]^2$, leading to a higher concentration of samples near the origin.
  \end{itemize}
\end{enumerate}

The combination yields four scenarios: Fixed Uniform, Fixed Clustered, Random Uniform, and Random Clustered. A total of 10\% of the sites are allocated for training and validation (with an 80/20 split, respectively), while the remaining 90\% are reserved for testing. Ten replications are conducted for each scenario-model combination. The performance of the estimated quantiles is assessed using the following criteria:}}

\paragraph{Probabilistic Accuracy.}
We use the Continuous Ranked Probability Score (CRPS) to evaluate distributional forecasts. The population CRPS for true value $y$ and predicted CDF $\hat{F}$ is
\begin{equation}
    \text{CRPS}(y, \hat{F}) = \int_{-\infty}^\infty \big(\hat{F}(z) - \mathbf{1}_{y \leq z}\big)^2 dz.
\end{equation}
In our setting we only have quantile predictions $\hat{Y}_{\tau_k}$ at levels $\tau_1,\ldots,\tau_K$ (e.g., $\tau\in\{0.05,0.25,0.5,0.75,0.95\}$), not the full $\hat{F}$. We therefore \textbf{estimate} CRPS using a weighted quantile approximation (trapezoidal integration in $\tau$-space):
\begin{equation}
    \widehat{\text{CRPS}}(y, \{\hat{Y}_{\tau_k}\}) = 2 \sum_{k=1}^{K} \tilde{w}_k \, \rho_{\tau_k}(y - \hat{Y}_{\tau_k}),
\end{equation}
where $\rho_\tau(u) = u(\tau - \mathbf{1}_{u<0})$ is the check loss, and $\tilde{w}_k$ are trapezoidal weights from the quantile grid:
\[
\tilde{w}_1=\frac{\tau_2-\tau_1}{2},\quad
\tilde{w}_k=\frac{\tau_{k+1}-\tau_{k-1}}{2}\ (2\le k\le K-1),\quad
\tilde{w}_K=\frac{\tau_K-\tau_{K-1}}{2}.
\]
This is more accurate than uniform weights when quantile levels are not equally spaced. All CRPS values reported in this paper are the mean of $\widehat{\text{CRPS}}$ over the test set.
Intuitively, CRPS is an integral over the quantile level $\tau$, so each quantile should contribute according to its interval width in $\tau$-space; trapezoidal weights respect this geometry, whereas uniform weights can over-emphasize densely sampled $\tau$ regions and underweight sparse ones.

\paragraph{Distributional Forecasting Results}
Table~\ref{tab:crps-latest} reports CRPS on 2b-8 with 10 replicates. DA-STDK-MQ  improves CRPS in three scenarios, with the largest gain in the random-clustered setting, while showing a small degradation in the random-uniform setting.

\begin{table}[h]
\centering
\caption{CRPS comparison on 2b-8 (mean (SE), 10 replicates).}
\begin{tabular}{lcc}
\toprule
Scenario & STDK & DA-STDK-MQ \\
\midrule
Fixed, uniform & 0.1867 (0.0006) & 0.1862 (0.0004) \\
Fixed, clustered & 0.2329 (0.0019) & 0.2224 (0.0013) \\
Random, uniform & 0.1962 (0.0004) & 0.1984 (0.0007) \\
Random, clustered & 0.2814 (0.0009) & 0.2528 (0.0010) \\
\bottomrule
\end{tabular}
\label{tab:crps-latest}
\end{table}

\paragraph{Coverage Evaluation}

To evaluate the performance of the proposed method, we compare \textbf{STDKGC CQR } ({STDKGC}, STDK quantiles with a single conformal $\hat{q}_{\text{global}}$), which applies a single conformal adjustment $\hat{q}_{\text{global}}$ to all quantile outputs ~\cite{romano2019cqr}, with \textbf{Ours}, which denotes the proposed DA-STDK-{MQ}  equipped with cluster-aware CQR (DA-STDK-{MQ}  + cluster-aware CQR).

 The comparison in the coverage performance is conducted using the metrics PICP, QICE, and worst-10\% coverage. We assess empirical coverage of the 90\% prediction interval on the test set via the Prediction Interval Coverage Probability (PICP)~\cite{yao2019quality}:
\begin{equation}
    \text{PICP} = \frac{1}{N}\sum_{n=1}^{N} \mathbf{1}_{y_n \geq \hat{y}_n^{\text{low}}} \cdot \mathbf{1}_{y_n \leq \hat{y}_n^{\text{high}}},
\end{equation}
where $\hat{y}_n^{\text{low}}$ and $\hat{y}_n^{\text{high}}$ are the lower and upper bounds of the 90\% PI for the $n$-th test point. To measure uniformity of coverage across quantile bins, we use Quantile Interval Coverage Error (QICE)~\cite{han2022card} with $M=4$ intervals defined by consecutive predicted quantiles $\tau\in\{0.05,0.25,0.5,0.75,0.95\}$ (intervals $[\hat{Y}_{0.05},\hat{Y}_{0.25}]$, $[\hat{Y}_{0.25},\hat{Y}_{0.5}]$, $[\hat{Y}_{0.5},\hat{Y}_{0.75}]$, $[\hat{Y}_{0.75},\hat{Y}_{0.95}]$). The target proportion per interval is $1/M = 1/4$. Let $r_m$ be the empirical proportion of test points whose true value falls in the $m$-th interval; then
\begin{equation}
    \text{QICE} = \frac{1}{M}\sum_{m=1}^{M} \left| r_m - \frac{1}{M} \right|, \quad r_m = \frac{1}{N}\sum_{n=1}^{N} \mathbf{1}_{y_n \in \text{$m$-th QI}}.
\end{equation}
Lower QICE indicates more uniform coverage across the distribution. To diagnose spatial heterogeneity and under/over-coverage, we also report worst-10\% site coverage by the mean coverage of the 10\% of sites with lowest coverage (tail behavior).

 The results correspond to \textbf{STDKGC} versus \textbf{Ours} (cluster-aware CQR), respectively, in  Table~\ref{tab:coverage-latest}.  Since the nominal coverage level is 90\%, we evaluate PICP in terms of its closeness to 90\%. While \textbf{STDKGC} has a slightly better performance under uniform sampling, the difference is not large. In contrast, \textbf{Ours} shows much closer coverage level under clustered sampling (fixed-clustered and random-clustered). The improvement is particularly evident in the Worst10 metric, which reflects lower-tail site-wise coverage robustness. Under clustered sampling, \textbf{Ours} substantially reduces the Worst10 coverage deficit. These improvements highlight the limitations of a single global conformal adjustment in cluster settings and demonstrate the advantage of the proposed cluster-aware calibration.

\begin{figure}[t]
    \centering
    \includegraphics[width=0.98\textwidth]{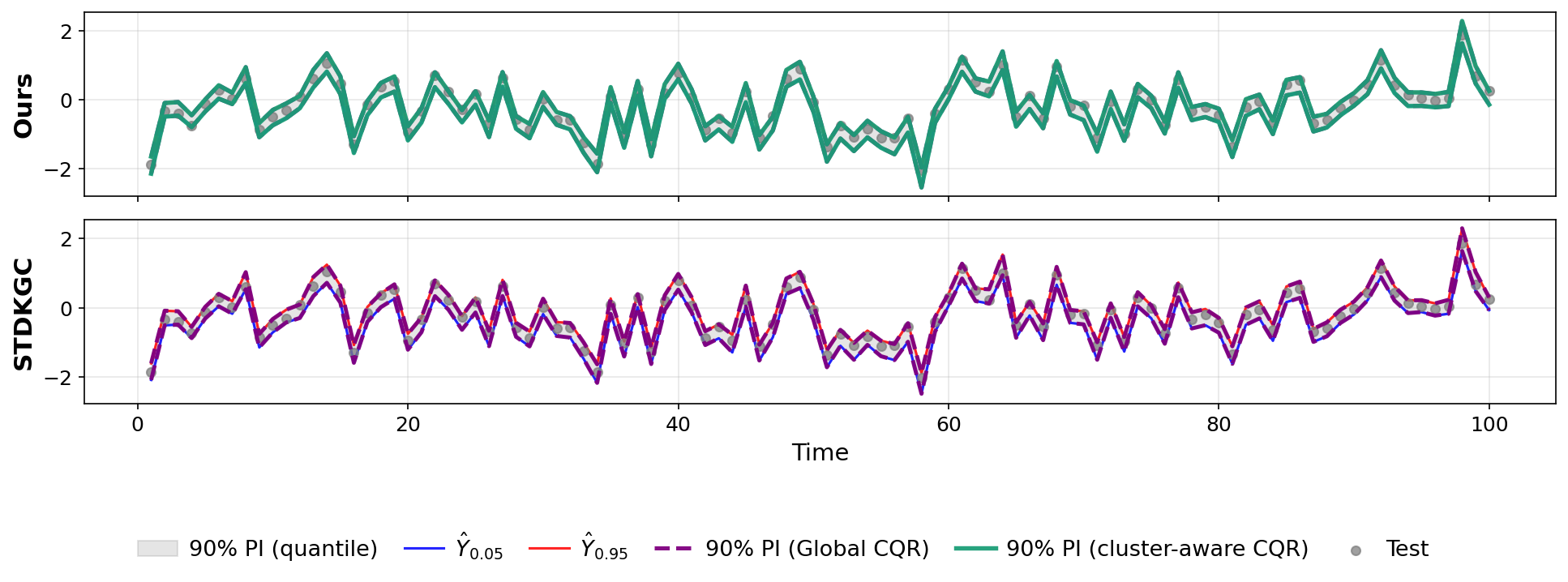}\\[-0.1em]
    {\small (a) Fixed-uniform}\\[0.4em]
    \includegraphics[width=0.98\textwidth]{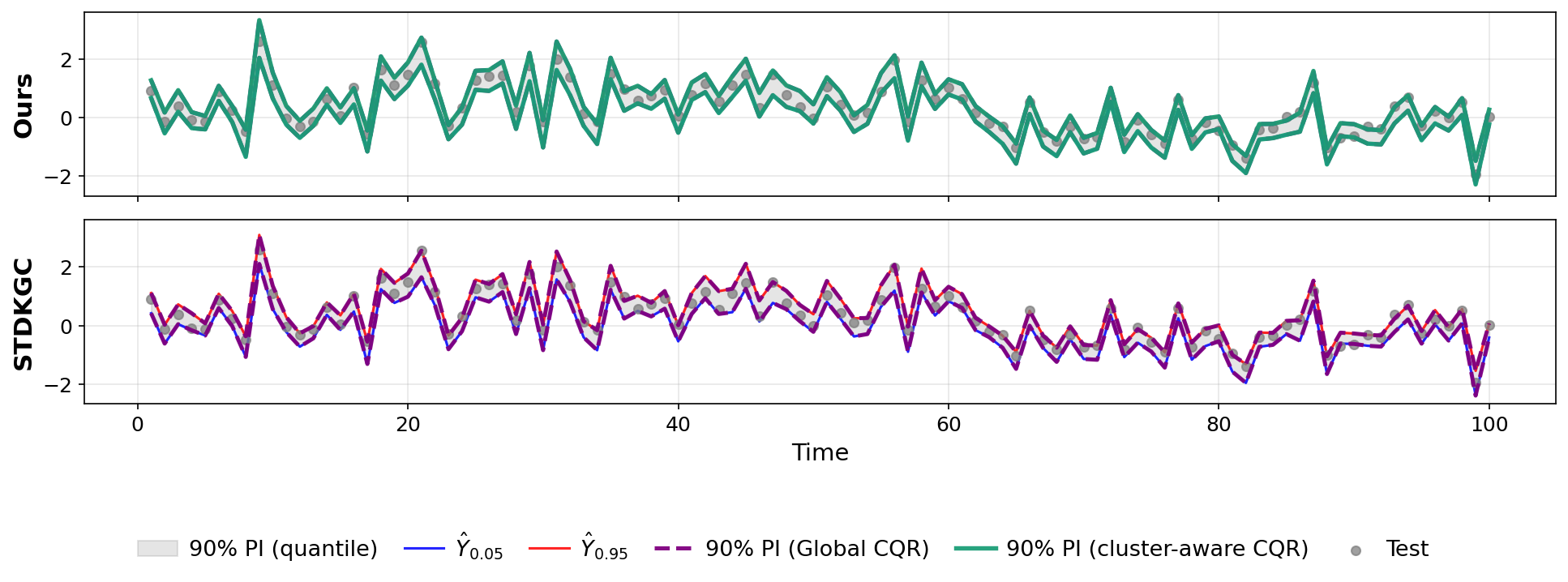}\\[-0.1em]
    {\small (b) Fixed-clustered}
    \caption{Time series at one representative test site for fixed-observation scenarios. \textbf{(a) Fixed-uniform}: For the site 23 at $(0.002,\, 0.236)$, the per-site empirical coverages of the 90\% PI are $0.72$ and $0.96$ for STDKGC and Ours, respectively. \textbf{(b) Fixed-clustered}: For the site 9971 at $(0.994,\, 0.717)$, the per-site empirical coverages of 90\% PI are $0.55$ and $0.93$ for STDKGC and  Ours, respectively. In each panel the \emph{top} subplot shows \textbf{Ours} (DA-STDK-MQ + cluster-aware CQR) and the \emph{bottom} shows \textbf{STDKGC} (baseline).}
    \label{fig:temporal-conformal-fixed}
\end{figure}

\begin{figure}[t]
    \centering
    \includegraphics[width=0.98\textwidth]{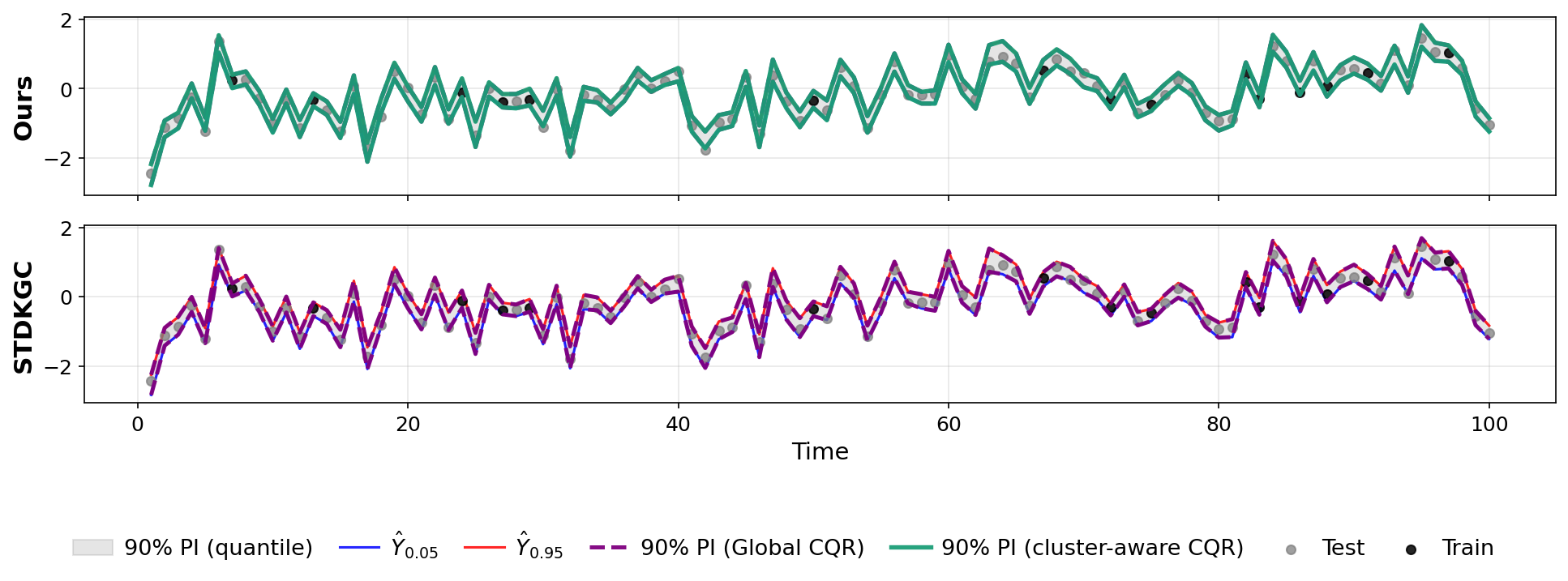}\\[-0.1em]
    {\small (a) Random-uniform}\\[0.4em]
    \includegraphics[width=0.98\textwidth]{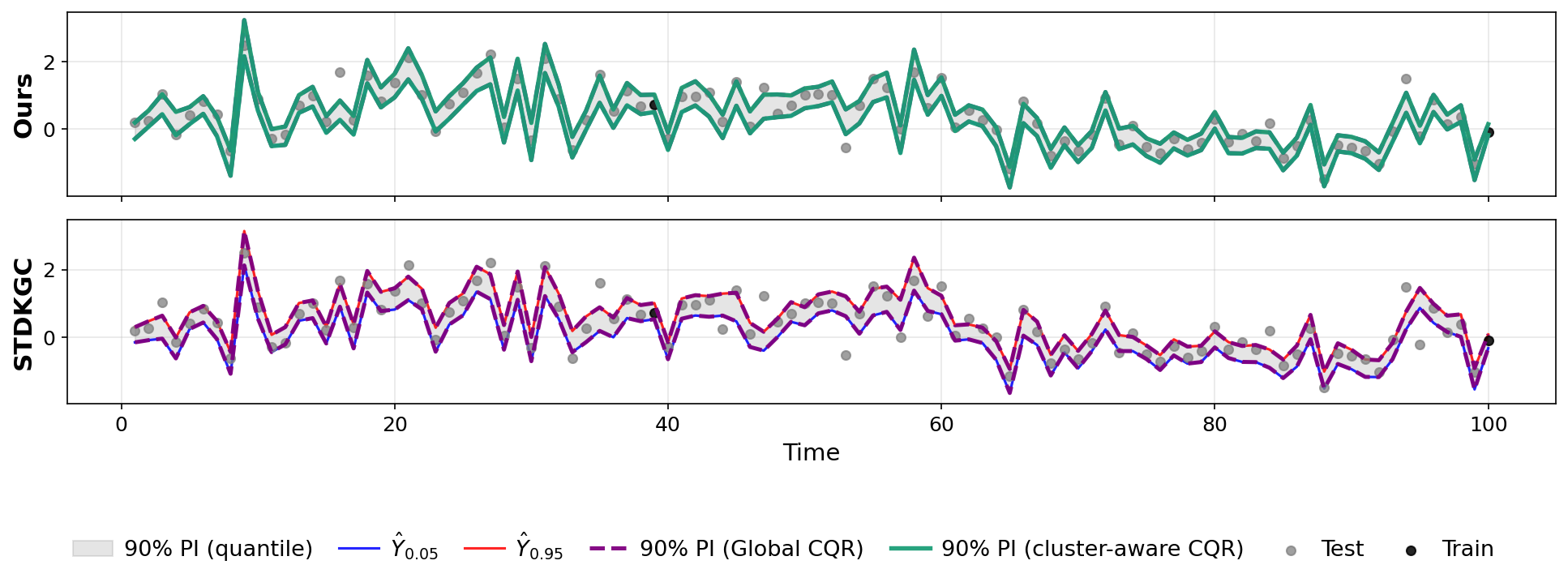}\\[-0.1em]
    {\small (b) Random-clustered}
    \caption{Random-observation analogue of Figure~\ref{fig:temporal-conformal-fixed}. Sites are chosen where \textbf{Ours} achieves coverage close to the 90\% nominal level while \textbf{STDKGC} is notably miscalibrated. \textbf{(a) Random-uniform}: For site 181 at $(0.018,\, 0.813)$, the per-site empirical coverages of the 90\% PI are $1.00$ and $0.94$ for STDKGC and Ours, respectively. \textbf{(b) Random-clustered}: For site 9891 at $(0.988,\, 0.918)$, the per-site empirical coverages of the 90\% PI are $0.69$ and $0.86$ for STDKGC and Ours, respectively.}
    \label{fig:temporal-conformal-random}
\end{figure}

In addition,  \textbf{Ours} improves distributional calibration, yielding lower QICE values across all scenarios. 
Figures~\ref{fig:temporal-conformal-fixed} and~\ref{fig:temporal-conformal-random} show time series at one representative test site per scenario, split into fixed-observation and random-observation pairs for readability. Each panel contains two stacked subplots evaluated at the same site: the \emph{top} subplot is \textbf{Ours} (DA-STDK-MQ  + cluster-aware CQR) and the \emph{bottom} subplot is \textbf{STDKGC} (baseline). {Test locations with the largest empirical coverage gaps between \textbf{STDKGC} and \textbf{Ours} are selected, so that the comparison focuses on regions where cluster-aware calibration is most informative.} Common legend: gray band = nominal quantile interval; blue and red lines = $\hat{Y}_{0.05}$ and $\hat{Y}_{0.95}$ of the underlying model; purple dashed lines = 90\% PI expanded with $\hat q_{\mathrm{global}}$; green solid lines (top subplot only) = 90\% PI expanded with the cluster-aware $\hat q_c$ (Algorithm~\ref{alg:cluster-cqr}); gray and black dots = test and train values; each subplot title reports the per-site empirical coverage. In the clustered scenarios (Figure~\ref{fig:temporal-conformal-fixed}(b) and Figure~\ref{fig:temporal-conformal-random}(b)), the top subplot keeps the true trajectory inside the 90\% PI more often than the bottom subplot, consistent with Table~\ref{tab:coverage-latest}.

\begin{sidewaystable}[htbp]
\centering
\caption{PICP, QICE, and worst-10\% site coverage on the latest run (mean (SE), 10 replicates): STDKGC vs Ours (DA-STDK-MQ  + cluster-aware CQR).}
\begin{tabular}{lcccccc}
\toprule
Scenario & PICP (STDKGC) & PICP (Ours) & QICE (STDKGC) & QICE (Ours) & Worst10 (STDKGC) & Worst10 (Ours) \\
\midrule
Fixed Uniform & 92.4\% (0.07) & 93.1\% (0.08) & 0.0429 (0.0004) & 0.0401 (0.0003) & 79.0\% (0.38) & 79.0\% (0.43) \\
Fixed Clustered & 86.9\% (0.22) & 92.1\% (0.16) & 0.0331 (0.0005) & 0.0318 (0.0002) & 63.5\% (0.66) & 73.9\% (0.46) \\
Random Uniform & 90.9\% (0.04) & 91.5\% (0.03) & 0.0390 (0.0002) & 0.0359 (0.0003) & 83.7\% (0.07) & 84.7\% (0.06) \\
Random Clustered & 80.9\% (0.13) & 90.7\% (0.10) & 0.0478 (0.0003) & 0.0386 (0.0003) & 62.4\% (0.23) & 81.0\% (0.28) \\
\bottomrule
\end{tabular}
\label{tab:coverage-latest}
\end{sidewaystable}

\begin{figure}[p]
  \centering
  \begin{subfigure}[b]{0.32\textwidth}
    \centering
    \includegraphics[width=\linewidth]{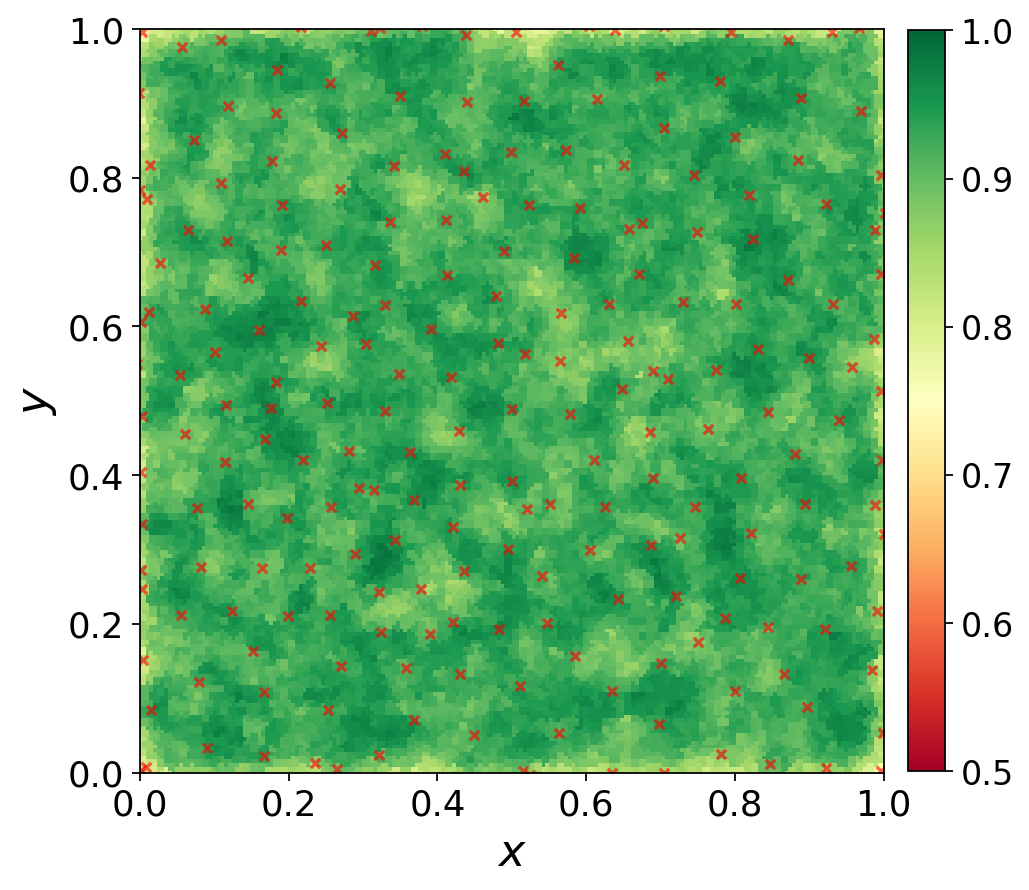}
    \caption{STDKGC}
    \label{fig:scov-fu-stdkgc}
  \end{subfigure}\hfill
  \begin{subfigure}[b]{0.32\textwidth}
    \centering
    \includegraphics[width=\linewidth]{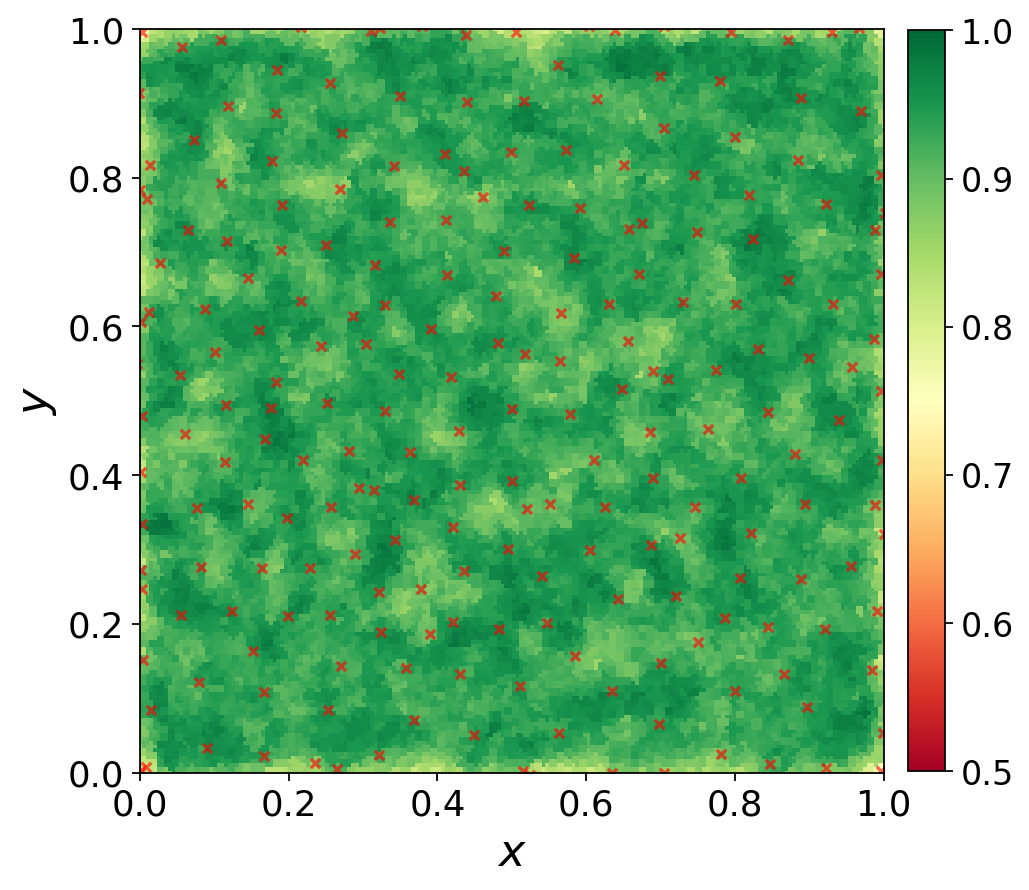}
    \caption{Ours}
    \label{fig:scov-fu-ours}
  \end{subfigure}\hfill
  \begin{subfigure}[b]{0.34\textwidth}
    \centering
    \includegraphics[width=\linewidth]{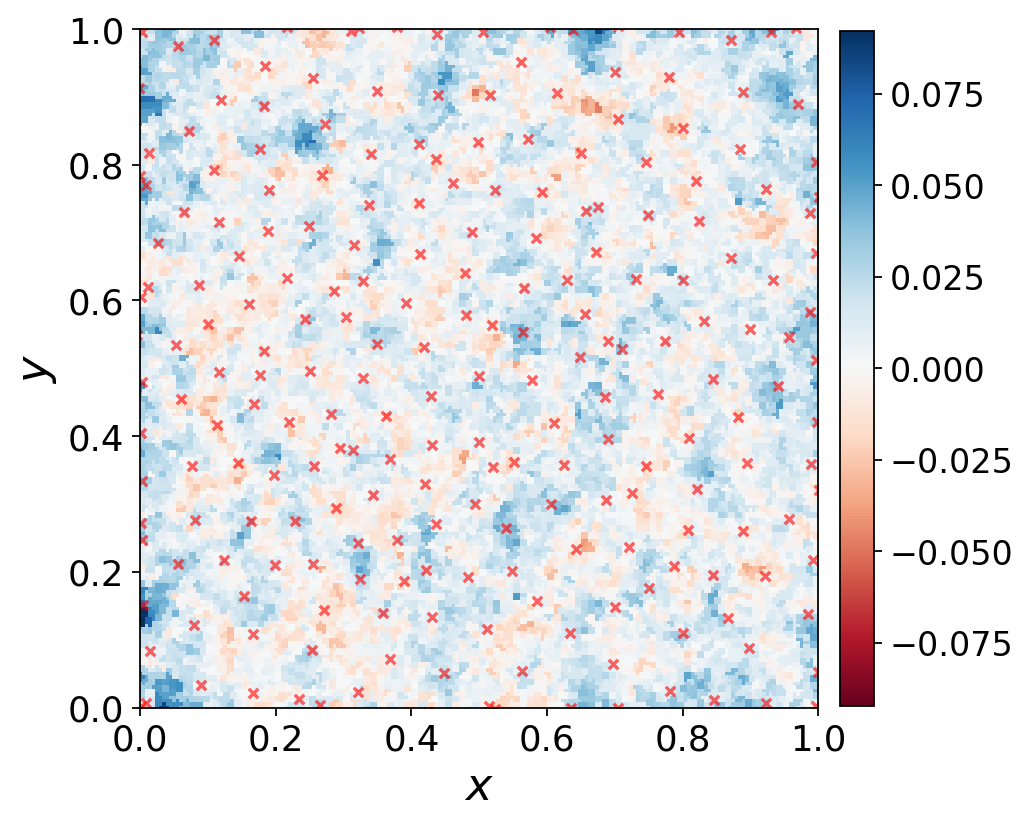}
    \caption{$\Delta = \text{Ours} - \text{STDKGC}$}
    \label{fig:scov-fu-delta}
  \end{subfigure}
  \\[0.6em]
  {\small (a) Fixed-uniform}\\[0.4em]
  \begin{subfigure}[b]{0.32\textwidth}
    \centering
    \includegraphics[width=\linewidth]{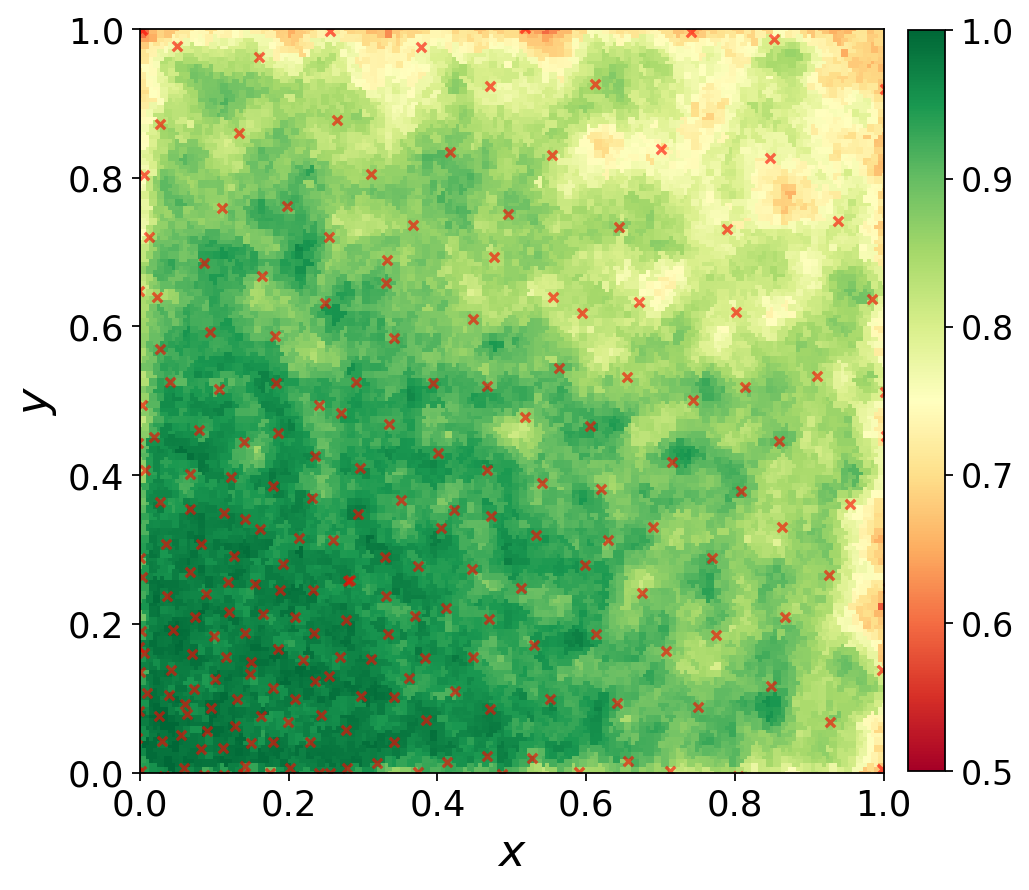}
    \caption{STDKGC}
    \label{fig:scov-fc-stdkgc}
  \end{subfigure}\hfill
  \begin{subfigure}[b]{0.32\textwidth}
    \centering
    \includegraphics[width=\linewidth]{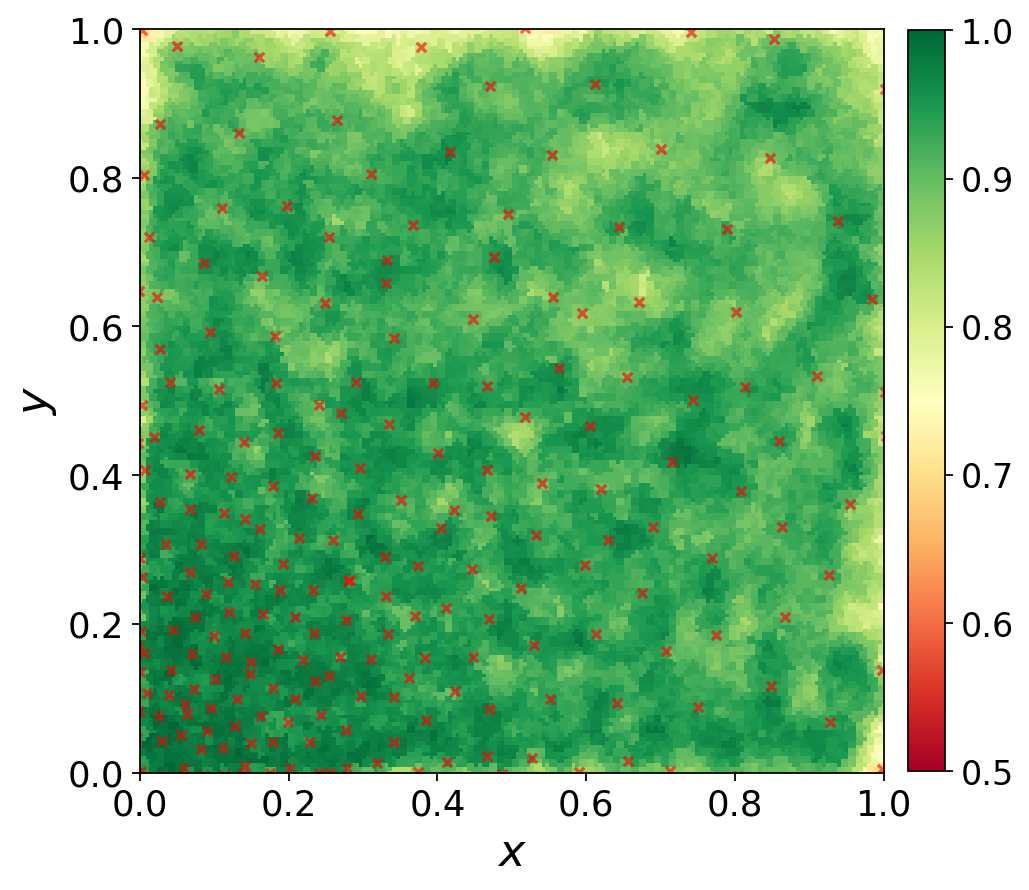}
    \caption{Ours}
    \label{fig:scov-fc-ours}
  \end{subfigure}\hfill
  \begin{subfigure}[b]{0.32\textwidth}
    \centering
    \includegraphics[width=\linewidth]{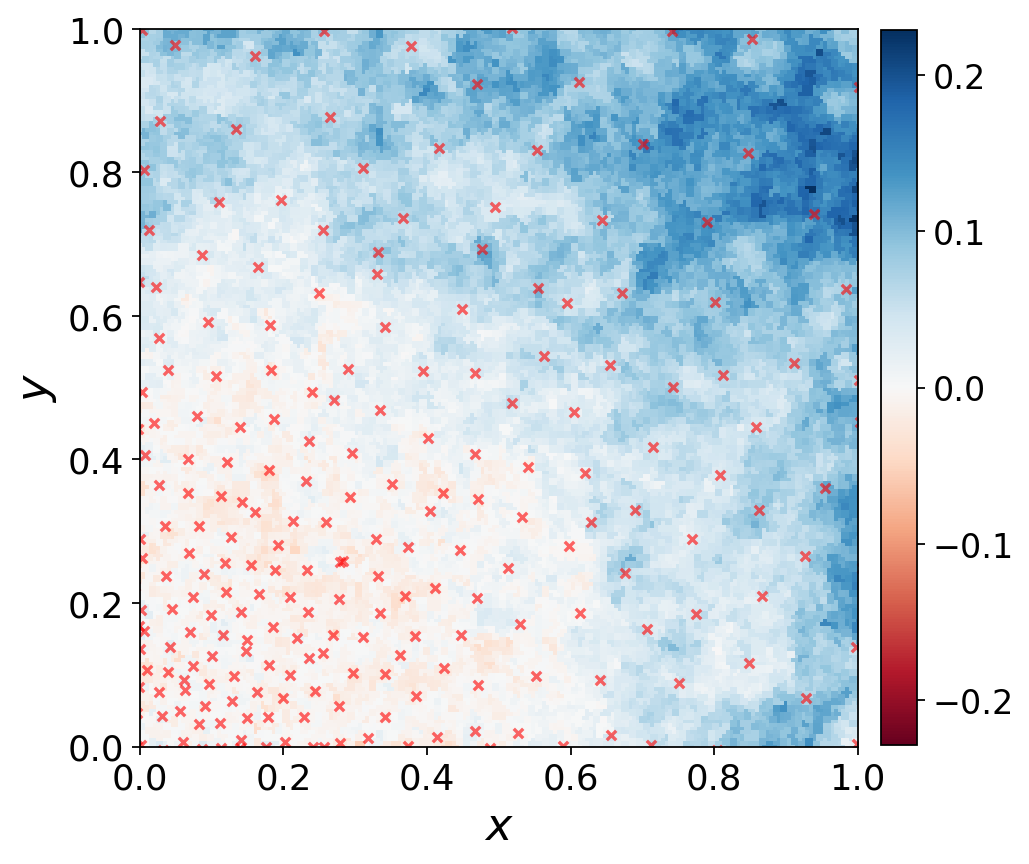}
    \caption{$\Delta = \text{Ours} - \text{STDKGC}$}
    \label{fig:scov-fc-delta}
  \end{subfigure}
  \\[0.4em]
  {\small (b) Fixed-clustered}
  \caption{Spatial coverage comparison for fixed-observation scenarios: (a) Fixed-uniform (top row); (b) Fixed-clustered (bottom row). Each row contains three sub-panels. \textbf{STDKGC}: per-site empirical 90\%-PI coverage of STDK + global CQR. \textbf{Ours}: per-site empirical 90\%-PI coverage of DA-STDK-MQ + cluster-aware CQR. $\boldsymbol{\Delta = \text{Ours} - \text{STDKGC}}$: site-wise difference. In the STDKGC and Ours sub-panels, green $=$ coverage at or above the 90\% target and red $=$ under-coverage. In the $\Delta$ sub-panel, blue $=$ Ours improves over STDKGC and red $=$ the opposite. Red $\times$ markers $=$ learned spatial basis centers. In Fixed-clustered, $\Delta$ is broadly positive across most of the domain.}  
  \label{fig:spatial-coverage-fixed}
\end{figure}

\begin{figure}[p]
  \centering
  \begin{subfigure}[b]{0.32\textwidth}
    \centering
    \includegraphics[width=\linewidth]{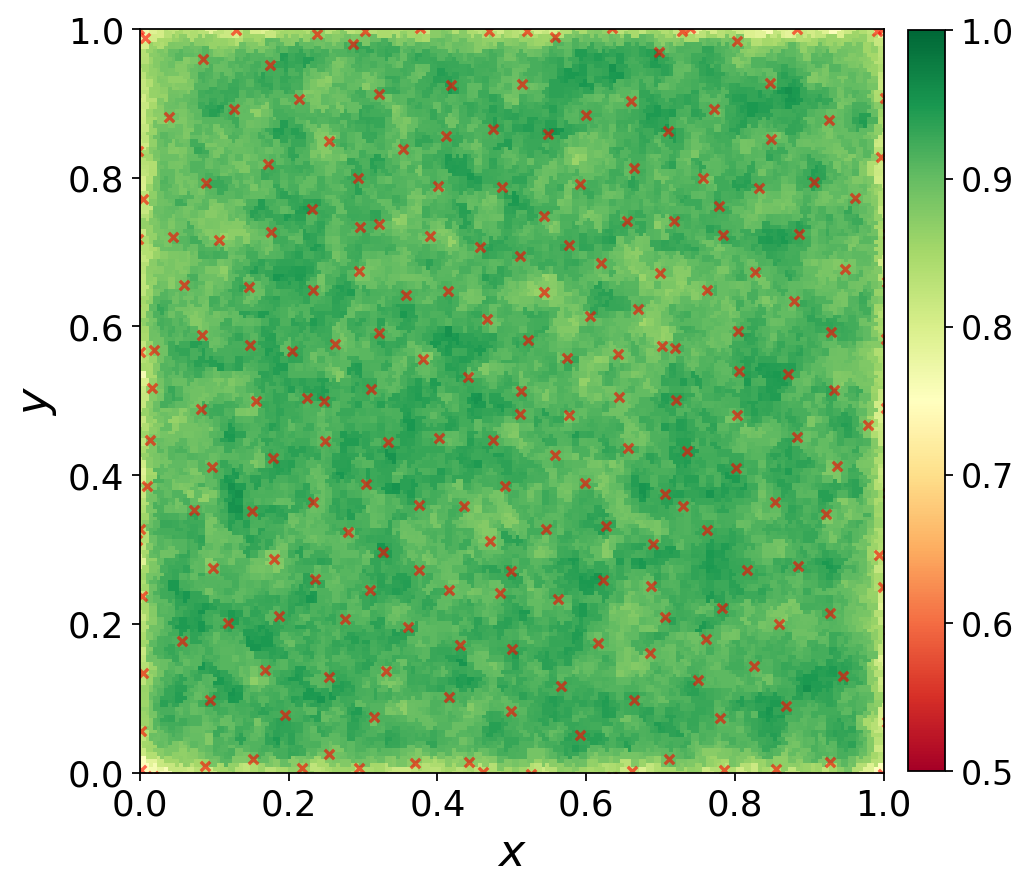}
    \caption{STDKGC}
  \end{subfigure}\hfill
  \begin{subfigure}[b]{0.32\textwidth}
    \centering
    \includegraphics[width=\linewidth]{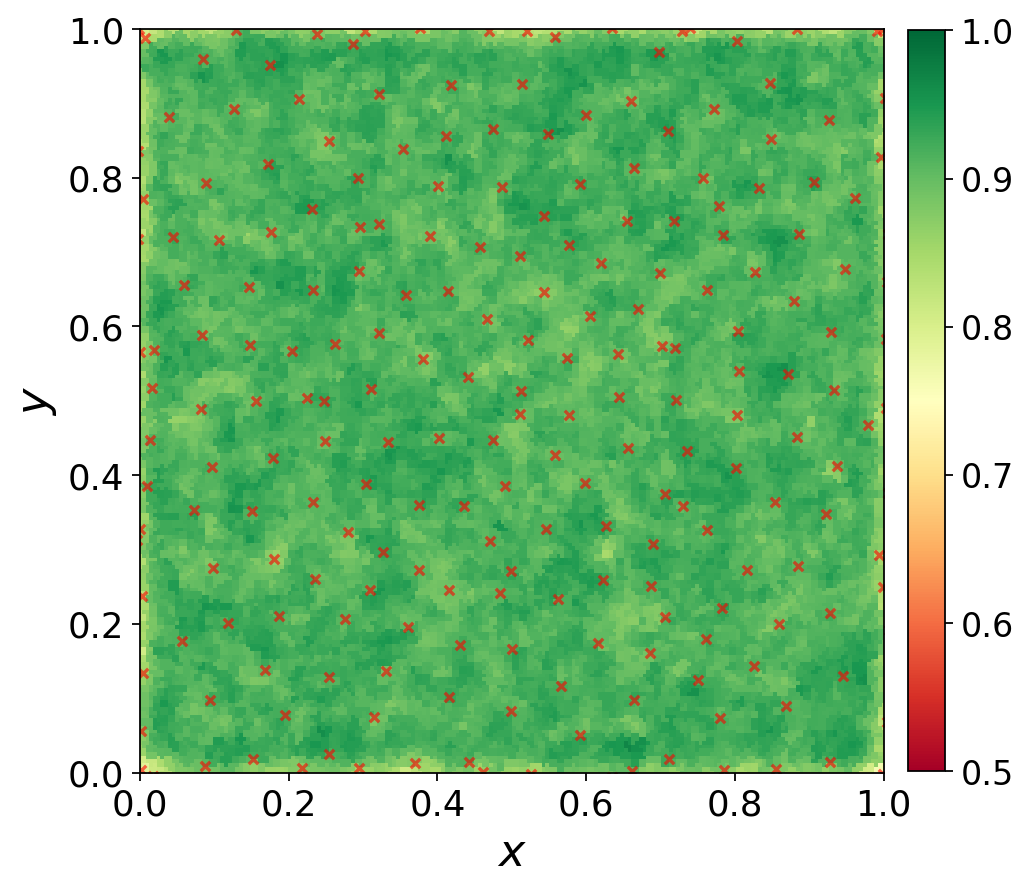}
    \caption{Ours}
  \end{subfigure}\hfill
  \begin{subfigure}[b]{0.34\textwidth}
    \centering
    \includegraphics[width=\linewidth]{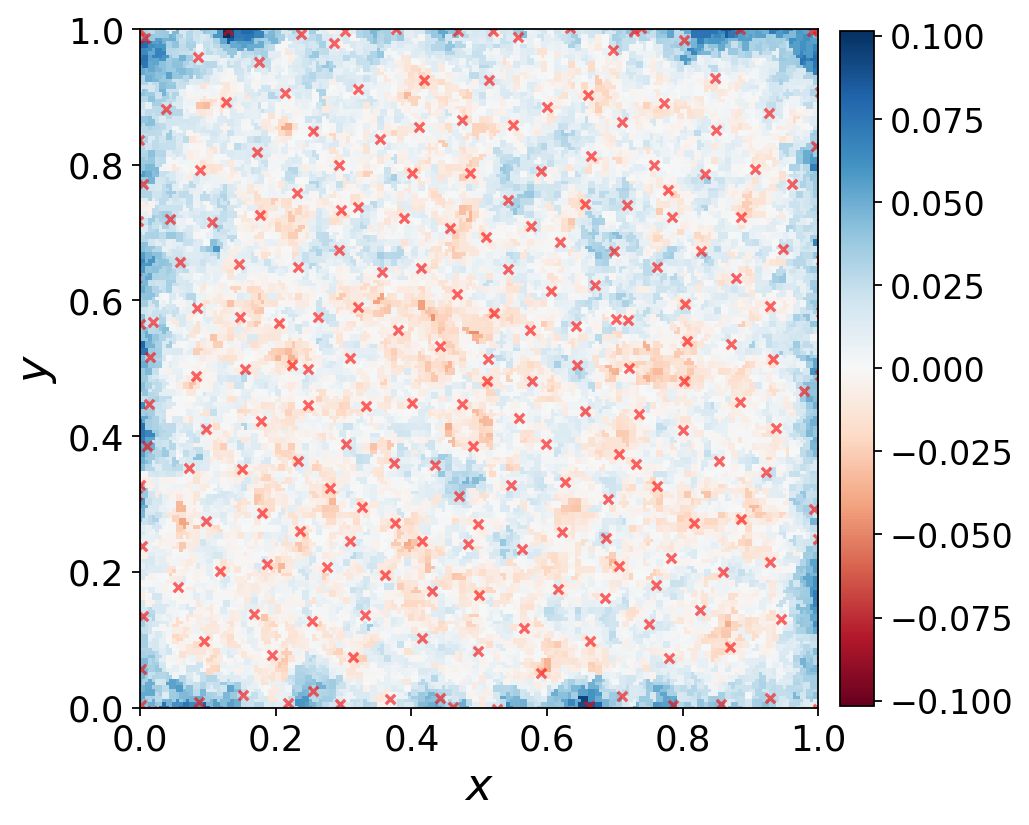}
    \caption{$\Delta = \text{Ours} - \text{STDKGC}$}
  \end{subfigure}
  \\[0.6em]
  {\small (a) Random-uniform}\\[0.4em]
  \begin{subfigure}[b]{0.32\textwidth}
    \centering
    \includegraphics[width=\linewidth]{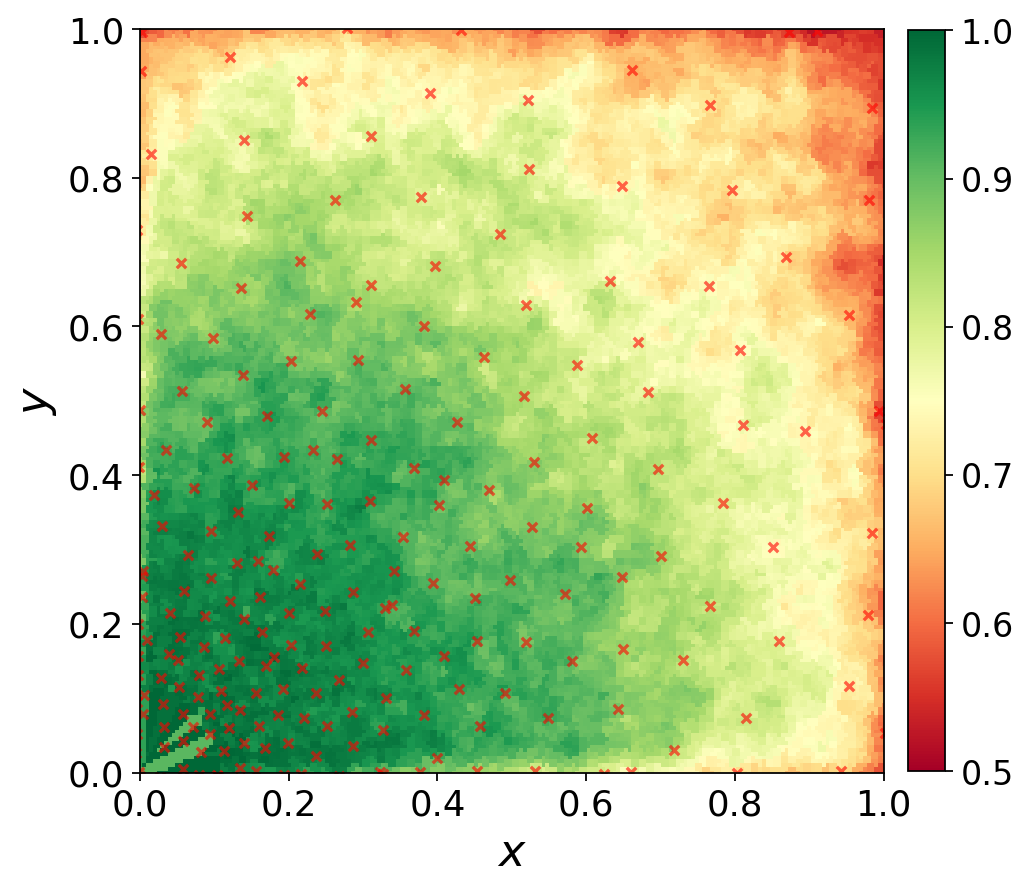}
    \caption{STDKGC}
  \end{subfigure}\hfill
  \begin{subfigure}[b]{0.32\textwidth}
    \centering
    \includegraphics[width=\linewidth]{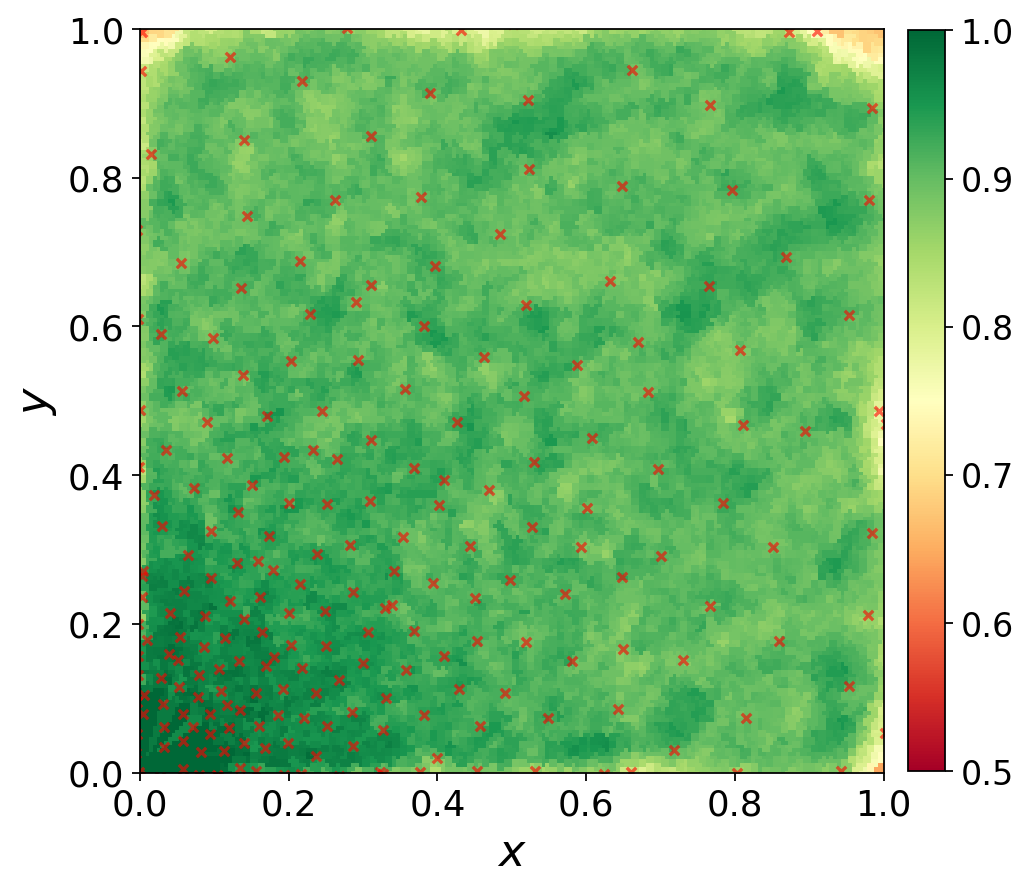}
    \caption{Ours}
  \end{subfigure}\hfill
  \begin{subfigure}[b]{0.32\textwidth}
    \centering
    \includegraphics[width=\linewidth]{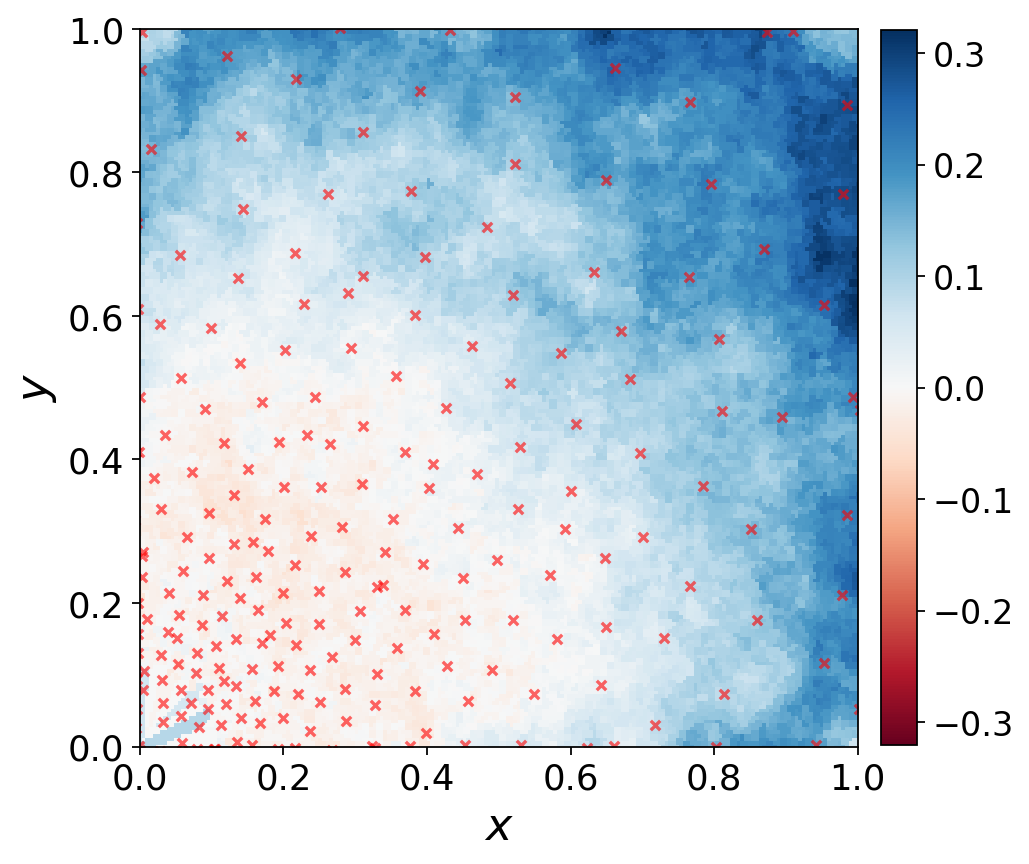}
    \caption{$\Delta = \text{Ours} - \text{STDKGC}$}
  \end{subfigure}
  \\[0.4em]
  {\small (b) Random-clustered}
  \caption{Spatial coverage comparison for random-observation scenarios: (a) Random-uniform; (b) Random-clustered. Same three-sub-panel layout and colour conventions as Figure~\ref{fig:spatial-coverage-fixed}. Red $\times$ markers $=$ spatial basis centers. Under Random-clustered, $\Delta$ is again predominantly positive across the domain, consistent with Table~\ref{tab:coverage-latest}.}
  \label{fig:spatial-coverage-random}
\end{figure}

We further compare the spatial coverage patterns between \textbf{STDKGC} and \textbf{Ours} under fixed-observation and random-observation scenarios, as shown in Figures~\ref{fig:spatial-coverage-fixed} and~\ref{fig:spatial-coverage-random}, respectively.
Each figure presents three panels per row: (1) coverage of \textbf{STDKGC}, (2) coverage of \textbf{Ours}, and (3) the difference $\Delta = \textbf{Ours} - \textbf{STDKGC}$. Columns correspond to the uniform and clustered observation structures.
In panels (1)--(2), green indicates higher empirical coverage and red indicates lower coverage relative to the target level (90\%). In panel (3), blue represents positive $\Delta$ and red represents negative $\Delta$. Red $\times$ markers denote the spatial basis centers.
Under uniform observations, $\Delta$ is close to zero and spatially scattered. In contrast, under clustered observations, $\Delta$ is predominantly positive across broad regions, indicating that \textbf{Ours} achieves superior coverage performance, consistent with the PICP and Worst10 results reported in Table~\ref{tab:coverage-latest}. 

\subsection{Real Data Application}



\begin{figure}[p]
\centering

\includegraphics[width=0.48\textwidth ]{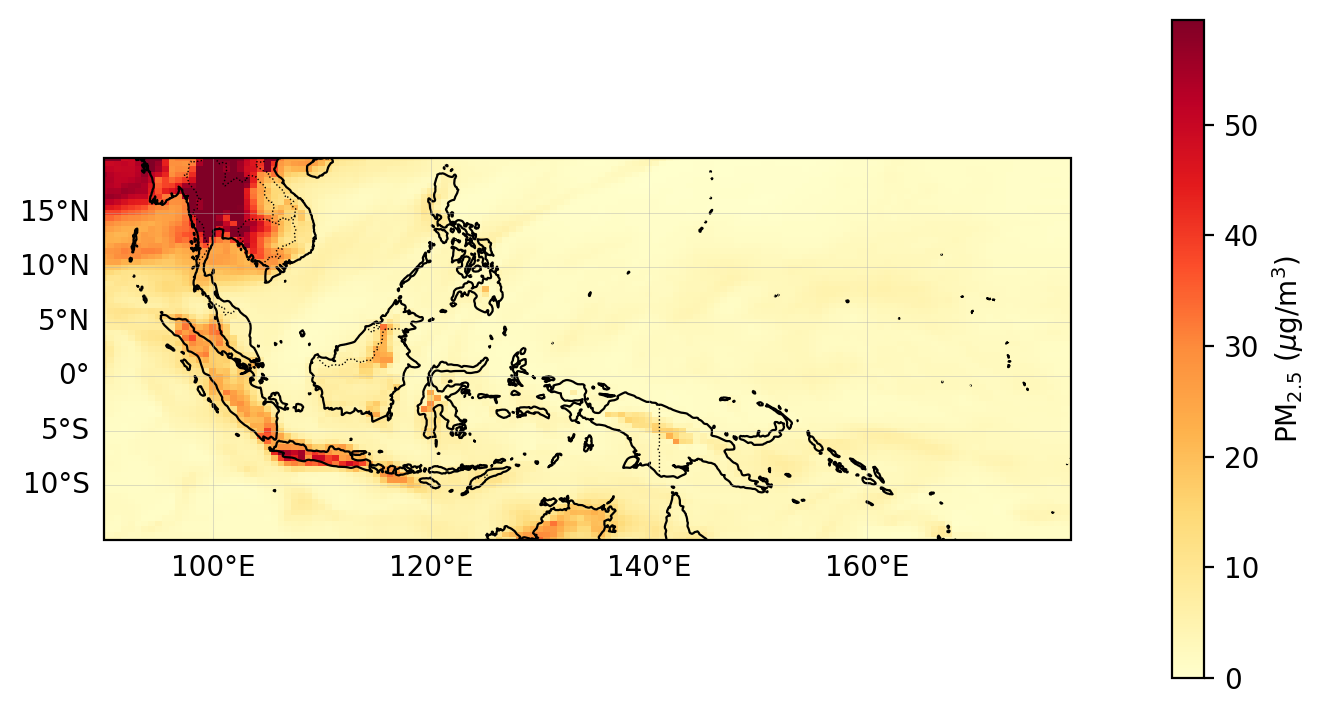}\hfill
\includegraphics[width=0.48\textwidth]{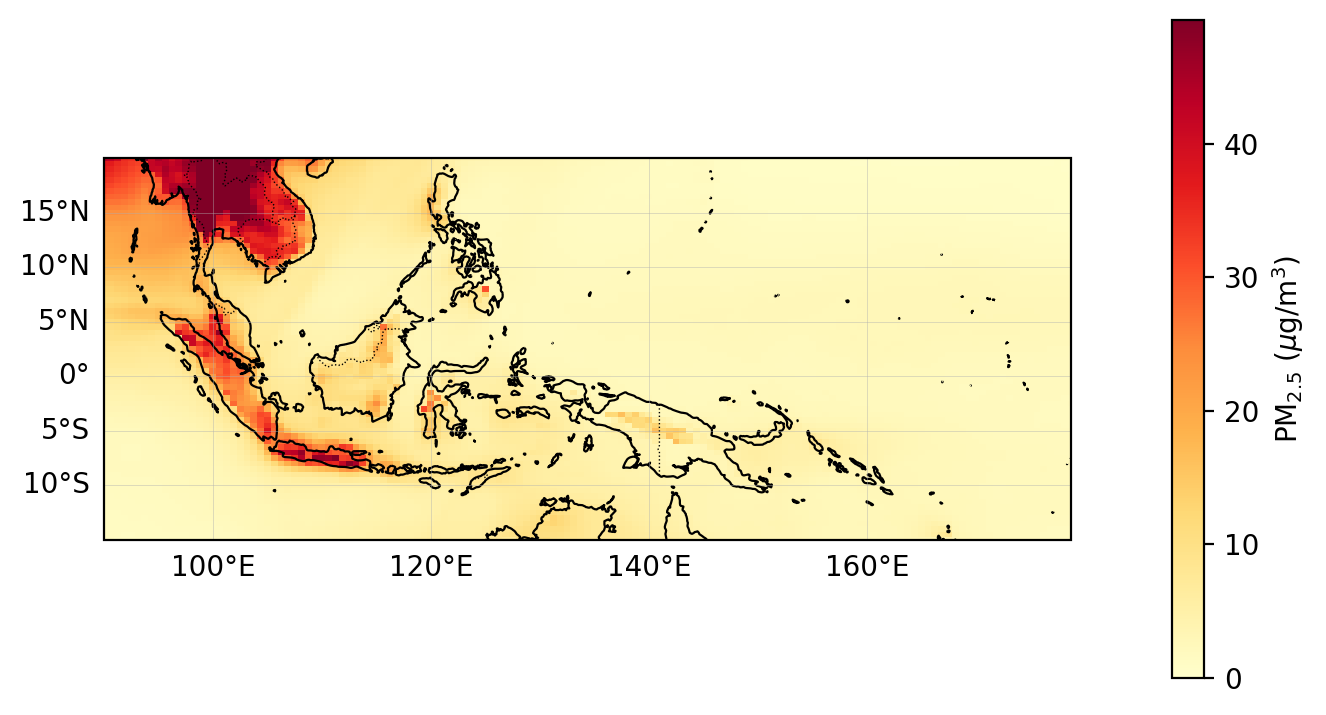}\\[0.5cm]

\includegraphics[width=0.48\textwidth]{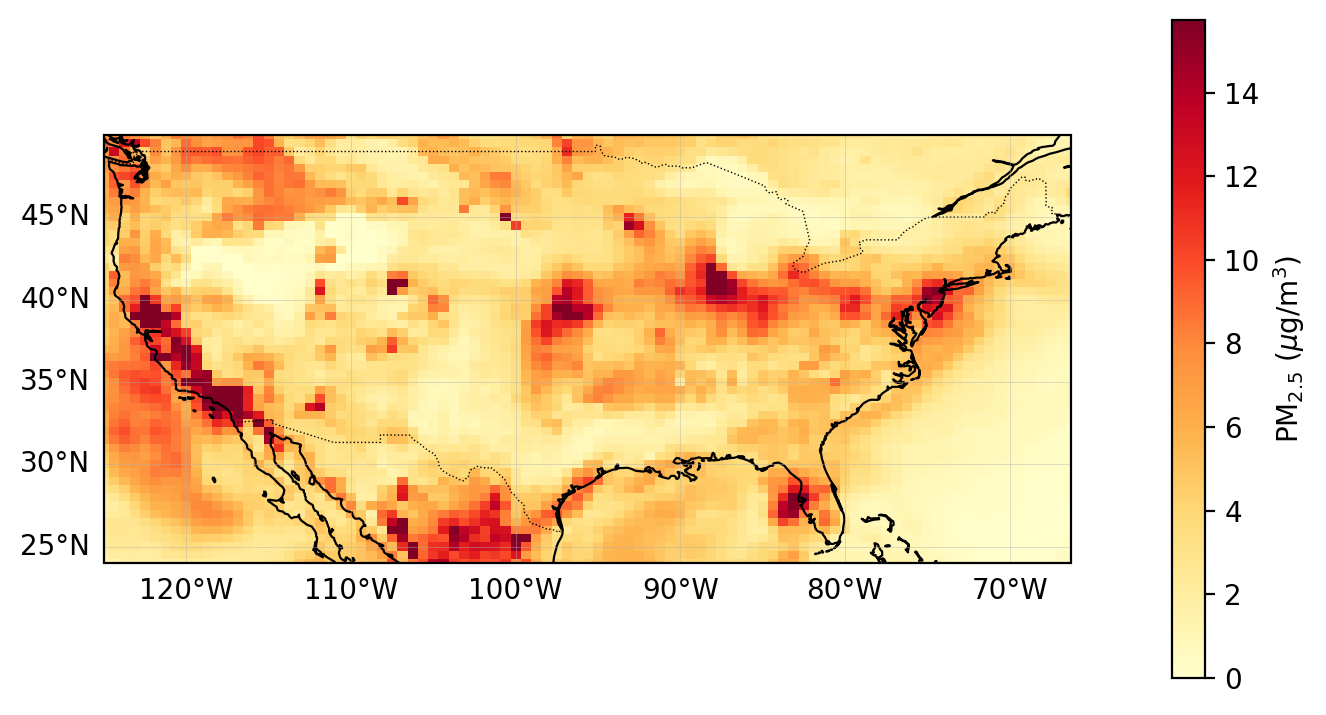}\hfill
\includegraphics[width=0.48\textwidth]{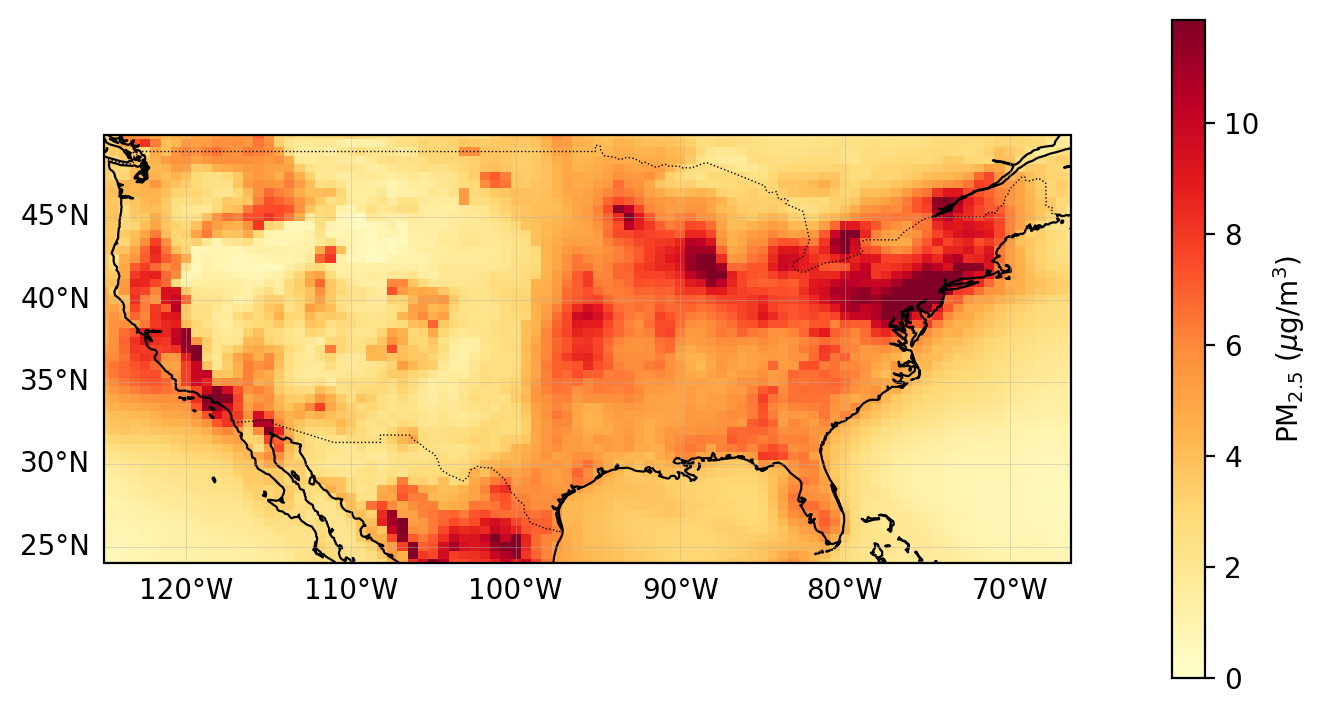}\\[0.5cm]

\includegraphics[width=0.48\textwidth]{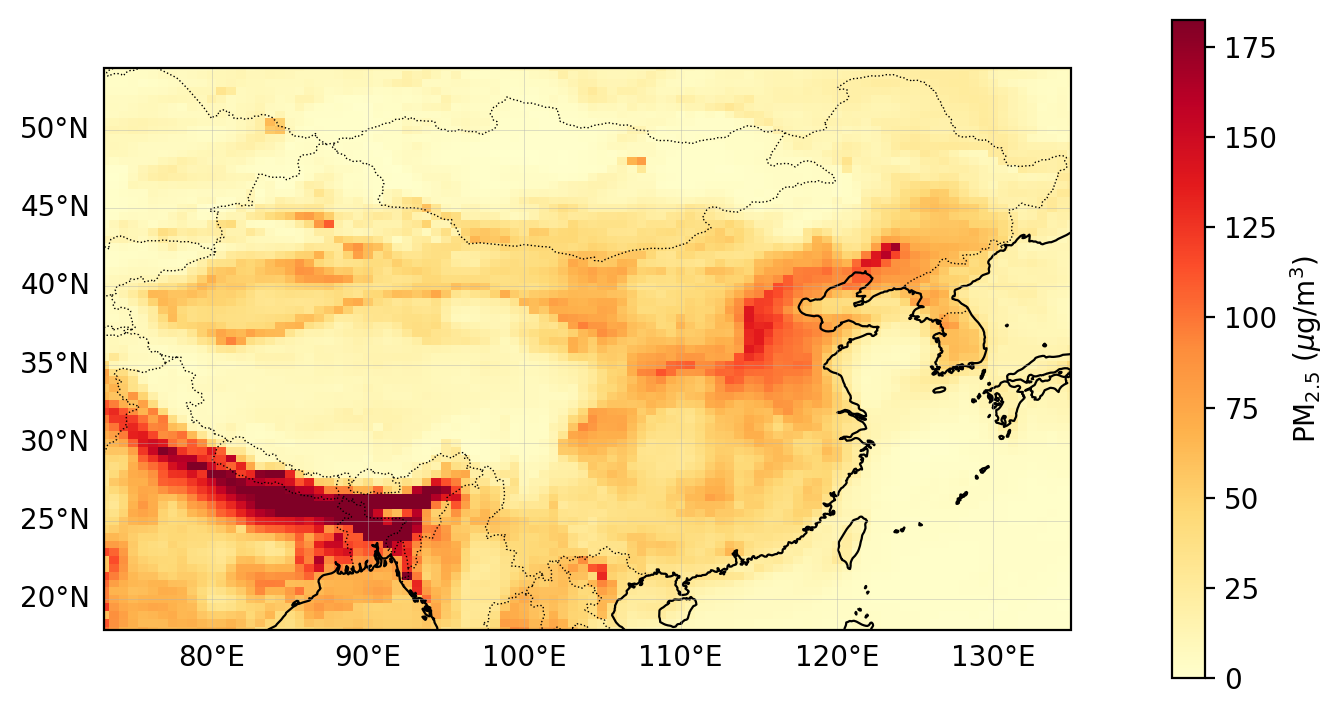}\hfill
\includegraphics[width=0.48\textwidth]{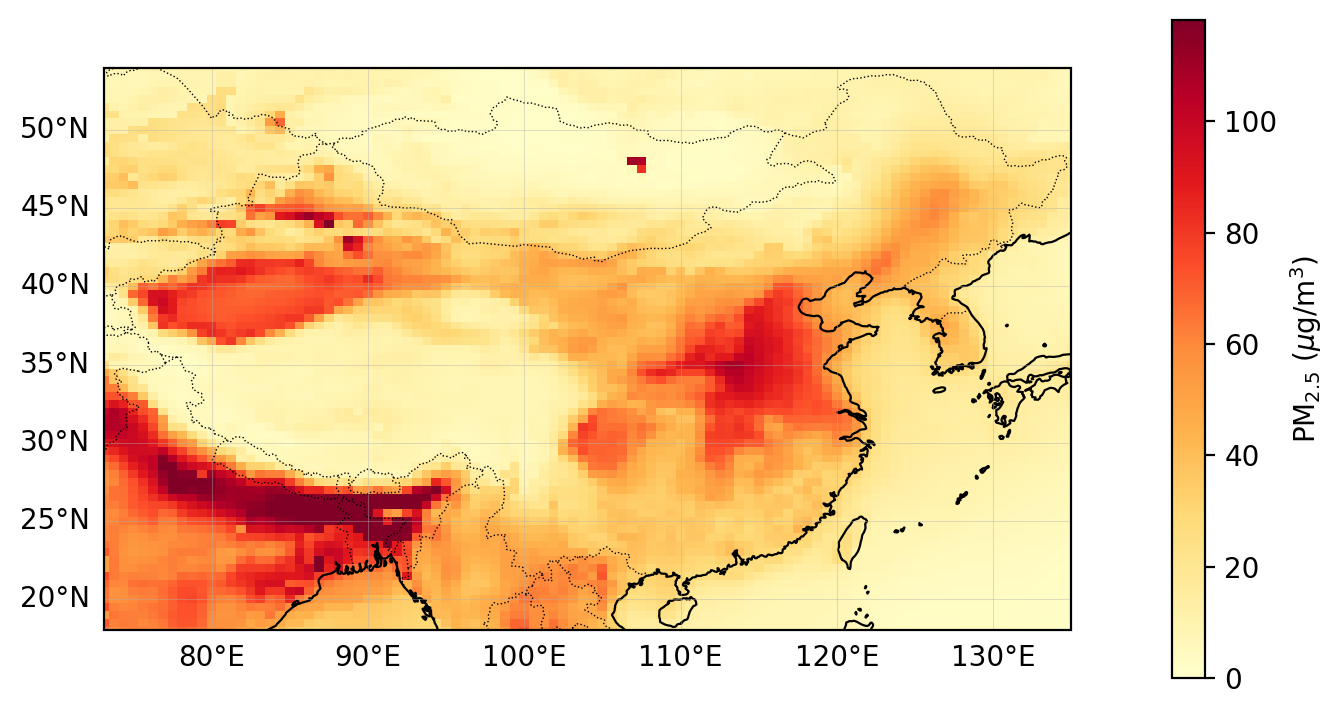}\\[-0.1em]
\caption{MERRA2 PM${2.5}$ concentration over the three real-data regions. Rows (top to bottom): Banda Sea, United States, China. Columns (left to right): Daily mean at 2020-02-19 and Temporal mean over $T = 100$ days (2020-01-01 to 2020-04-09).}
\label{fig:temporal-conformal-scenarios-pm25}
\end{figure}

The MERRA2 CNN HAQAST PM\textsubscript{2.5} dataset provides satellite-derived daily mean particulate matter concentrations on a $0.5\degree \times 0.625\degree$ latitude--longitude grid. We consider PM\textsubscript{2.5} datasets from three geographic regions. One-day realizations and their corresponding mean fields from a single replication for the regions of interest are presented in Figure~\ref{fig:temporal-conformal-scenarios-pm25}. 

\begin{itemize}
  \item \textbf{Banda Sea (Indonesia):} $10\degree$S--$2\degree$S latitude, $118\degree$E--$132\degree$E longitude. A tropical maritime region with relatively homogeneous PM\textsubscript{2.5} levels. For the Banda Sea dataset, the high concentrations of PM\textsubscript{2.5} occurred over land areas.\\

  \item \textbf{United States:} $24\degree$N--$50\degree$N latitude, $125\degree$W--$66\degree$W longitude. A large continental domain with diverse emission sources and meteorological conditions.\\
  \item \textbf{China:} $18\degree$N--$54\degree$N latitude, $73\degree$E--$135\degree$E longitude. A region with strong spatial gradients in PM\textsubscript{2.5} due to industrial activity. 
\end{itemize}

For each region, the data are restricted to $T = 100$ timesteps, corresponding to the period from 2020-01-01 to 2020-04-09 and $S \leq 10{,}000$ grid cells.
In these datasets, the PM\textsubscript{2.5} data uses real geographic coordinates, enabling the spatial kernel to capture actual geographic distance relationships. Similar to the simulation study, we consider four observation settings: Fixed Uniform, Fixed Clustered, Random Uniform, and Random Clustered. The performance of the proposed method is evaluated using \textbf{CRPS}, \textbf{PICP} and \textbf{QICE}, as defined in the simulation study.

\subsection{Results}

In each table, we report the mean and standard deviation (in parentheses) across 10 replications using different random seeds for the train-test splits, consistent with the procedure used for the KAUST dataset. A total of 10\% of the sites are allocated for training and validation (with an 80/20 split, respectively), while the remaining 90\% are reserved for testing.

 The ``Improvement (\%)'' column is defined so that positive values indicate better performance of \textbf{Ours} relative to \textbf{STDKGC}. 
For CRPS and QICE, where  smaller values indicate better performance, the improvement is computed as
$100 \times \frac{\text{STDKGC} - \text{Ours}}{\text{STDKGC}}.$
For PICP, since all reported coverage values are below the nominal level of 90\%, larger PICP values indicate coverage closer to the nominal level. Therefore, the improvement is computed as
$100 \times \frac{\text{Ours} - \text{STDKGC}}{\text{STDKGC}}.$
Thus, positive values indicate that \textbf{Ours} improves the corresponding metric, whereas negative values indicate deterioration.


\subsubsection{PM\textsubscript{2.5} --- Banda Sea}

\begin{table}[htbp]
\centering
\caption{Results for PM\textsubscript{2.5} Banda Sea. Mean $\pm$ std over 10 replications.}
\begin{tabular}{l l S[table-format=1.4] @{${}\pm{}$} S[table-format=0.4] S[table-format=1.4] @{${}\pm{}$} S[table-format=0.4] r}
\toprule
Scenario & Metric & \multicolumn{2}{c}{{\textbf{STDKGC}}} & \multicolumn{2}{c}{\textbf{Ours}}& Improvement (\%)\\
\midrule
\multirow{3}{*}{Fixed Uniform}
 & CRPS & 1.0981 & 0.0297 & 1.1005 & 0.0287 & -0.2\% \\
 & PICP & 0.8514 & 0.0032 & 0.8499 & 0.0033 & -$0.2$\% \\
 & QICE & 0.0371 & 0.0008 & 0.0375 & 0.0008 & -1.0\% \\
\midrule
\multirow{3}{*}{Fixed Clustered}
 & CRPS & 1.2316 & 0.0659 & 1.2270 & 0.0706 & +$0.4$\% \\
 & PICP & 0.8241 & 0.0039 & 0.8329 & 0.0039 & +1.1\% \\
 & QICE & 0.0440 & 0.0010 & 0.0418 & 0.0010 & +$5.0$\% \\
\midrule
\multirow{3}{*}{Random Uniform}
 & CRPS & 0.9311 & 0.0043 & 0.9249 & 0.0044 & +$0.7$\% \\
 & PICP & 0.8753 & 0.0026 & 0.8759 & 0.0044 & +0.1\% \\
 & QICE & 0.0314 & 0.0007 & 0.0316 & 0.0005 & -0.8\% \\
\midrule
\multirow{3}{*}{Random Clustered}
 & CRPS & 1.0510 & 0.0069 & 1.0409 & 0.0105 & +$1.0$\% \\
 & PICP & 0.8308 & 0.0031 & 0.8583 & 0.0080 & +3.3\% \\
 & QICE & 0.0423 & 0.0008 & 0.0354 & 0.0020 & +$16.3$\% \\
\bottomrule
\end{tabular}
\label{tab:pm25-banda}
\end{table}

 CRPS ranges from 0.93 to 1.23.
Similar to the simulation studies, DA-STDK-MQ  improves primarily under clustered observation scenarios, with the largest gains in Random Clustered ($1.0\%$ CRPS, $3.3\%$ PICP, $16.3\%$ QICE).

\subsubsection{PM\textsubscript{2.5} --- United States}

\begin{table}[htbp]
\centering
\caption{Results for PM\textsubscript{2.5} United States. Mean $\pm$ std over 10 replications.}
\label{tab:pm25-us}
\begin{tabular}{l l S[table-format=1.4] @{${}\pm{}$} S[table-format=0.4] S[table-format=1.4] @{${}\pm{}$} S[table-format=0.4] r}
\toprule
Scenario & Metric & \multicolumn{2}{c}{{ \textbf{STDKGC}}} & \multicolumn{2}{c}{{\textbf{Ours}}} & Improvement (\%) \\
\midrule
\multirow{3}{*}{Fixed Uniform}
 & CRPS & 0.9993 & 0.0207 & 1.0018 & 0.0210 & -0.2\% \\
 & PICP & 0.7863 & 0.0054 & 0.7939 & 0.0114 & +1.0\% \\
 & QICE & 0.0534 & 0.0014 & 0.0515 & 0.0029 & +$3.5$\% \\
\midrule
\multirow{3}{*}{Fixed Clustered}
 & CRPS & 1.1969 & 0.0330 & 1.1591 & 0.0284 & +$3.2$\% \\
 & PICP & 0.7448 & 0.0168 & 0.7542 & 0.0120 & +1.3\% \\
 & QICE & 0.0638 & 0.0042 & 0.0614 & 0.0030 & +$3.7$\% \\
\midrule
\multirow{3}{*}{Random Uniform}
 & CRPS & 0.9625 & 0.0085 & 0.9699 & 0.0095 & -0.8\% \\
 & PICP & 0.8254 & 0.0146 & 0.8304 & 0.0095 & +0.6\% \\
 & QICE & 0.0437 & 0.0037 & 0.0424 & 0.0024 & +$2.8$\% \\
\midrule
\multirow{3}{*}{Random Clustered}
 & CRPS & 1.2150 & 0.0106 & 1.1395 & 0.0052 & +$6.2$\% \\
 & PICP & 0.7551 & 0.0187 & 0.8016 & 0.0092 & +6.2\% \\
 & QICE & 0.0612 & 0.0047 & 0.0496 & 0.0023 & +$19.0$\% \\
\bottomrule
\end{tabular}
\end{table}

The U.S.\ PM\textsubscript{2.5} results show the strongest DA-STDK-{MQ}  improvements among the PM\textsubscript{2.5} regions. Under the Random Clustered setting, DA-STDK-MQ reduces CRPS by 6.2\%, improves coverage by 6.2\%, and decreases QICE by 19.0\%. Similarly, under the Fixed Clustered setting, it achieves meaningful improvements, with CRPS, PICP, and QICE improved by 3.2\%, 1.3\%, and 3.7\%, respectively.


\subsubsection{PM\textsubscript{2.5} --- China}

\begin{table}[htbp]
\centering
\caption{Results for PM\textsubscript{2.5} China. Mean $\pm$ std over 10 replications.}
\begin{tabular}{l l S[table-format=1.3] @{${}\pm{}$} S[table-format=0.3] S[table-format=1.3] @{${}\pm{}$} S[table-format=0.3] r}
\toprule
Scenario & Metric & \multicolumn{2}{c}{{ \textbf{STDKGC}}} & \multicolumn{2}{c}{{\textbf{Ours}}}  & Improvement (\%)\\
\midrule
\multirow{3}{*}{Fixed Uniform}
 & CRPS & 5.574 & 0.123 & 5.593 & 0.103 & -0.3\% \\
 & PICP & 0.795 & 0.005 & 0.793 & 0.007 & -$0.3$\% \\
 & QICE & 0.051 & 0.001 & 0.052 & 0.002 & -1.1\% \\
\midrule
\multirow{3}{*}{Fixed Clustered}
 & CRPS & 6.181 & 0.158 & 6.020 & 0.127 & +$2.6$\% \\
 & PICP & 0.754 & 0.007 & 0.768 & 0.005 & +1.9\% \\
 & QICE & 0.062 & 0.002 & 0.058 & 0.001 & +$5.8$\% \\
\midrule
\multirow{3}{*}{Random Uniform}
 & CRPS & 4.756 & 0.025 & 4.752 & 0.029 & +$0.1$\% \\
 & PICP & 0.859 & 0.007 & 0.857 & 0.006 & $-0.2$\% \\
 & QICE & 0.035 & 0.002 & 0.036 & 0.002 & -1.0\% \\
\midrule
\multirow{3}{*}{Random Clustered}
 & CRPS & 5.442 & 0.028 & 5.204 & 0.034 & +$4.4$\% \\
 & PICP & 0.789 & 0.008 & 0.826 & 0.005 & +4.7\% \\
 & QICE & 0.053 & 0.002 & 0.043 & 0.001 & +$17.5$\% \\
\bottomrule
\end{tabular}
\label{tab:pm25-china}
\end{table}
\newpage
China PM\textsubscript{2.5} has the highest CRPS values among the three regions, reflecting stronger spatial gradients in pollution levels. DA-STDK-{MQ}  achieves meaningful improvements under both clustered scenarios: $2.6\%$ and $4.4\%$ CRPS for Fixed and Random Clustered, respectively.


Table~\ref{tab:cross-crps}–\ref{tab:cross-qice} summarize the improvement (\%) of CRPS, QICE and PICP of DA-STDK-{MQ}  over STDKGC across all three datasets and four observation scenarios. They demonstrate that the proposed method outperforms well for geographically clustered data.

\begin{table}[htbp]
\centering
\caption{\textbf{Improvement of CRPS}}
\begin{tabular}{l rrrr}
\toprule
Dataset & Fixed Unif. & Fixed Clust. & Rand.\ Unif. & Rand.\ Clust. \\
\midrule
PM2.5 Banda Sea       & -0.2  & +$0.4$  & +$0.7$  & +$1.0$  \\
PM2.5 U.S.            & -0.2  & +$3.2$  & -0.8  & +$6.2$  \\
PM2.5 China           & -0.3  & +$2.6$  & +$0.1$  & +$4.4$  \\
\bottomrule
\end{tabular}
\label{tab:cross-crps}
\end{table}

\begin{table}[htbp]
\centering
\caption{{\textbf{Improvement of PICP}}}
\begin{tabular}{l rrrr}
\toprule
Dataset & {Fixed Unif.} & {Fixed Clust.} & {Rand.\ Unif.} & {Rand.\ Clust.} \\
\midrule
{PM2.5 Banda Sea} & {$-0.2$} & {+1.1} & {+0.1} & {+3.3} \\
{PM2.5 U.S.} & {+1.0} & {+1.3} & {+0.6} & {+6.2} \\
{PM2.5 China} &{$-0.3$} & {+1.9} & {$-0.2$} & {+4.7} \\
\bottomrule
\end{tabular}
\label{tab:cross-coverage}
\end{table}

\begin{table}[htbp]
\centering
\caption{\textbf{Improvement of QICE}}
\begin{tabular}{l rrrr}
\toprule
Dataset & Fixed Unif. & Fixed Clust. & Rand.\ Unif. & Rand.\ Clust. \\
\midrule
PM2.5 Banda Sea       & -1.0  & +$5.0$  & -0.8  & +$16.3$ \\
PM2.5 U.S.            & +$3.5$  & +$3.7$  & +$2.8$  & +$19.0$ \\
PM2.5 China           & -1.1  & +$5.8$  & -1.0  & +$17.5$ \\
\bottomrule
\end{tabular}
\label{tab:cross-qice}
\end{table}

\section{Conclusion}
DeepKriging-style models, such as STDK, achieve scalability via basis-function embeddings and stochastic gradient learning; however, fixed regular-grid spatial bases remain inefficient under highly non-uniform sampling, often over-representing sparse regions while under-resolving dense clusters. In this work, we address this limitation through cluster-adaptive spatial bases and emphasize distributional forecasting with reliable uncertainty quantification. Specifically, we develop (i) a cluster-adaptive spatial basis with learnable centers and scales, initialized from the spatial sampling density and jointly optimized with network weights; (ii) a multi-quantile joint training framework with non-crossing regularization; and (iii) a cluster-aware conformal calibration layer that adjusts prediction-interval widths at the cluster level, with a global fallback for small clusters. Simulation studies and PM\textsubscript{2.5} analysis demonstrate that the proposed framework improves coverage accuracy and distributional forecasting under clustered data structures, compared with global conformal methods.

\bmhead{Acknowledgements}

Lim’s research was supported by National Research Foundation of Korea (NRF) grant funded by the Korea government (MSIT) (RS-2024-
00335033). Huang’s research  is partially supported  by NSTC 113-2118-M-259-001-MY2 and NDHU Funding 114T2560-03. Wu’s research  is partially supported  by NSTC 113-2118-M-259-002-MY2 and NDHU Funding 114T2560-03. Wang’s research is supported by NSTC113-2118-M005-005-MY2.

\newpage
\begin{appendices}
\section{Additional Results for the Remaining KAUST Competition Datasets}\label{secA1}
Our proposed methodology is further applied to additional KAUST competition datasets, including 2a-7, 2a-8, 2a-9, 2b-7, and 2b-9. Depending on the number of observation locations, we consider different basis-function settings: $(9,25,36)$ for $S=1000$ and $(25,81,121)$ for $S=10000$. All other simulation settings remain the same as those described in Section~\ref{sec:method:Simulation}. The corresponding results are presented and discussed in this section.

\begin{table}[h]
\centering
\caption{CRPS on KAUST 2a-7 ($S = 1{,}000$, basis $[9,25,36]$; mean (SE), 10 replicates).}
\begin{tabular}{l S[table-format=0.4] @{${}\;$} S[table-format=0.4] S[table-format=0.4] @{${}\;$} S[table-format=0.4]}
\toprule
Scenario & \multicolumn{2}{c}{STDK} & \multicolumn{2}{c}{DA-STDK-MQ } \\
\midrule
Fixed, uniform     & 0.6168 & {(0.0021)} & 0.6152 & {(0.0015)} \\
Fixed, clustered   & 0.6377 & {(0.0034)} & 0.6255 & {(0.0050)} \\
Random, uniform    & 0.5335 & {(0.0004)} & 0.5318 & {(0.0004)} \\
Random, clustered  & 0.5523 & {(0.0025)} & 0.5441 & {(0.0007)} \\
\bottomrule
\end{tabular}
\label{tab:crps-2a7}
\vskip0.5cm
\centering
\caption{CRPS on KAUST 2a-8 ($S = 1{,}000$, basis $[9,25,36]$; mean (SE), 10 replicates).}
\begin{tabular}{l S[table-format=0.4] @{${}\;$} S[table-format=0.4] S[table-format=0.4] @{${}\;$} S[table-format=0.4]}
\toprule
Scenario & \multicolumn{2}{c}{STDK} & \multicolumn{2}{c}{DA-STDK-MQ } \\
\midrule
Fixed, uniform     & 0.4498 & {(0.0024)} & 0.4547 & {(0.0029)} \\
Fixed, clustered   & 0.4659 & {(0.0027)} & 0.4650 & {(0.0023)} \\
Random, uniform    & 0.4690 & {(0.0012)} & 0.4722 & {(0.0013)} \\
Random, clustered  & 0.5263 & {(0.0043)} & 0.4944 & {(0.0016)} \\
\bottomrule
\end{tabular}
\label{tab:crps-2a8}
\vskip0.5cm
\centering
\caption{CRPS  on KAUST 2a-9 ($S = 1{,}000$, basis $[9,25,36]$; mean (SE), 10 replicates).}
\begin{tabular}{l S[table-format=0.4] @{${}\;$} S[table-format=0.4] S[table-format=0.4] @{${}\;$} S[table-format=0.4]}
\toprule
Scenario & \multicolumn{2}{c}{STDK} & \multicolumn{2}{c}{DA-STDK-{MQ} } \\
\midrule
Fixed, uniform     & 0.1210 & {(0.0019)} & 0.1253 & {(0.0020)} \\
Fixed, clustered   & 0.1669 & {(0.0044)} & 0.1579 & {(0.0034)} \\
Random, uniform    & 0.1533 & {(0.0022)} & 0.1681 & {(0.0016)} \\
Random, clustered  & 0.2722 & {(0.0024)} & 0.2322 & {(0.0023)} \\
\bottomrule
\end{tabular}
\label{tab:crps-2a9}
\vskip0.5cm
\centering
\caption{CRPS on KAUST 2b-7 ($S = 10{,}000$, basis $[25,81,121]$; mean (SE), 10 replicates).}
\label{tab:crps-2b7}
\begin{tabular}{l S[table-format=0.4] @{${}\;$} S[table-format=0.4] S[table-format=0.4] @{${}\;$} S[table-format=0.4]}
\toprule
Scenario & \multicolumn{2}{c}{STDK} & \multicolumn{2}{c}{DA-STDK-{MQ} } \\
\midrule
Fixed, uniform     & 0.4315 & {(0.0006)} & 0.4324 & {(0.0006)} \\
Fixed, clustered   & 0.4720 & {(0.0013)} & 0.4696 & {(0.0015)} \\
Random, uniform    & 0.4133 & {(0.0003)} & 0.4145 & {(0.0002)} \\
Random, clustered  & 0.4602 & {(0.0005)} & 0.4530 & {(0.0004)} \\
\bottomrule
\end{tabular}
\vskip0.5cm
\centering
\caption{CRPS on KAUST 2b-9 ($S = 10{,}000$, basis $[25,81,121]$; mean (SE), 10 replicates).}
\label{tab:crps-2b9}
\begin{tabular}{l S[table-format=0.4] @{${}\;$} S[table-format=0.4] S[table-format=0.4] @{${}\;$} S[table-format=0.4]}
\toprule
Scenario & \multicolumn{2}{c}{STDK} & \multicolumn{2}{c}{DA-STDK-{MQ} } \\
\midrule
Fixed, uniform     & 0.0482 & {(0.0002)} & 0.0474 & {(0.0001)} \\
Fixed, clustered   & 0.0580 & {(0.0006)} & 0.0566 & {(0.0004)} \\
Random, uniform    & 0.0484 & {(0.0001)} & 0.0476 & {(0.0001)} \\
Random, clustered  & 0.0683 & {(0.0003)} & 0.0609 & {(0.0001)} \\
\bottomrule
\end{tabular}

\end{table}

{\paragraph{CRPS.}
CRPS information are shown in Table \ref{tab:crps-2a7} - \ref{tab:crps-2b9}. Across all five datasets, DA-STDK-MQ  consistently outperforms STDK in the \emph{clustered} scenarios,
with the largest CRPS reductions in the random-clustered setting:
\begin{itemize}
  \item 2a-9: 0.2722 $\to$ 0.2322 ($+$14.7\%), the largest relative improvement.
  \item 2a-8: 0.5263 $\to$ 0.4944 ($+$6.1\%).
  \item 2b-9: 0.0683 $\to$ 0.0609 ($+$10.8\%).
  \item 2b-7: 0.4602 $\to$ 0.4530 ($+$1.6\%).
  \item 2a-7: 0.5523 $\to$ 0.5441 ($+$1.5\%).
\end{itemize}
In uniform scenarios, the two models perform comparably, with DA-STDK-{MQ}  showing a small degradation
on some datasets (e.g., 2a-9 random-uniform: 0.1533 $\to$ 0.1681).
This is consistent with the main paper's 2b-8 finding: the adaptive basis provides the most benefit
when observations are spatially clustered.
}

\begin{sidewaystable}[htbp]
\centering
\caption{PICP, QICE, and worst-10\% site coverage on KAUST 2a-7 ($S = 1{,}000$, basis $[9,25,36]$; mean (SE), 10 replicates): STDKGC vs.\ Ours.}
\label{tab:cov-2a7}
\begin{tabular}{l r@{$\;$}l r@{$\;$}l r@{$\;$}l r@{$\;$}l r@{$\;$}l r@{$\;$}l}
\toprule
Scenario & \multicolumn{2}{c}{PICP (STDKGC)} & \multicolumn{2}{c}{PICP (Ours)} & \multicolumn{2}{c}{QICE (STDKGC)} & \multicolumn{2}{c}{QICE (Ours)} & \multicolumn{2}{c}{W10 (STDKGC)} & \multicolumn{2}{c}{W10 (Ours)} \\
\midrule
Fixed, uniform    & 80.0\% & (0.32) & 80.8\% & (0.31) & 0.0727 & (0.0014) & 0.0709 & (0.0006) & 66.4\% & (0.72) & 67.0\% & (0.51) \\
Fixed, clustered  & 77.2\% & (0.58) & 81.0\% & (0.58) & 0.0812 & (0.0021) & 0.0708 & (0.0024) & 62.2\% & (0.81) & 66.4\% & (0.77) \\
Random, uniform   & 89.7\% & (0.20) & 90.8\% & (0.18) & 0.0293 & (0.0005) & 0.0300 & (0.0006) & 82.1\% & (0.28) & 82.5\% & (0.34) \\
Random, clustered & 87.7\% & (0.33) & 90.1\% & (0.20) & 0.0380 & (0.0019) & 0.0321 & (0.0005) & 76.9\% & (0.79) & 81.4\% & (0.23) \\
\bottomrule
\end{tabular}%
\hskip0.3cm
\centering
\caption{PICP, QICE, and worst-10\% site coverage on KAUST 2a-8 ($S = 1{,}000$, basis $[9,25,36]$; mean (SE), 10 replicates): STDKGC vs.\ Ours.}
\label{tab:cov-2a8}
\begin{tabular}{l r@{$\;$}l r@{$\;$}l r@{$\;$}l r@{$\;$}l r@{$\;$}l r@{$\;$}l}
\toprule
Scenario & \multicolumn{2}{c}{PICP (STDKGC)} & \multicolumn{2}{c}{PICP (Ours)} & \multicolumn{2}{c}{QICE (STDKGC)} & \multicolumn{2}{c}{QICE (Ours)} & \multicolumn{2}{c}{W10 (STDKGC)} & \multicolumn{2}{c}{W10 (Ours)} \\
\midrule
Fixed, uniform    & 90.9\% & (0.24) & 91.5\% & (0.22) & 0.0398 & (0.0020) & 0.0376 & (0.0019) & 81.6\% & (0.64) & 82.1\% & (0.59) \\
Fixed, clustered  & 87.0\% & (0.42) & 90.0\% & (0.34) & 0.0615 & (0.0028) & 0.0532 & (0.0019) & 73.5\% & (0.79) & 77.7\% & (0.64) \\
Random, uniform   & 89.6\% & (0.14) & 90.5\% & (0.10) & 0.0446 & (0.0017) & 0.0421 & (0.0018) & 82.3\% & (0.30) & 82.3\% & (0.27) \\
Random, clustered & 82.6\% & (0.61) & 88.3\% & (0.25) & 0.0790 & (0.0065) & 0.0505 & (0.0024) & 68.1\% & (1.01) & 79.7\% & (0.38) \\
\bottomrule
\end{tabular}%
\hskip0.3cm
\centering
\caption{PICP, QICE, and worst-10\% site coverage on KAUST 2a-9 ($S = 1{,}000$, basis $[9,25,36]$; mean (SE), 10 replicates): STDKGC vs.\ Ours.}
\label{tab:cov-2a9}
\begin{tabular}{l r@{$\;$}l r@{$\;$}l r@{$\;$}l r@{$\;$}l r@{$\;$}l r@{$\;$}l}
\toprule
Scenario & \multicolumn{2}{c}{PICP (STDKGC)} & \multicolumn{2}{c}{PICP (Ours)} & \multicolumn{2}{c}{QICE (STDKGC)} & \multicolumn{2}{c}{QICE (Ours)} & \multicolumn{2}{c}{W10 (STDKGC)} & \multicolumn{2}{c}{W10 (Ours)} \\
\midrule
Fixed, uniform    & 90.5\% & (0.62) & 91.1\% & (0.48) & 0.0346 & (0.0007) & 0.0351 & (0.0006) & 69.3\% & (2.76) & 71.7\% & (2.02) \\
Fixed, clustered  & 84.0\% & (1.05) & 86.2\% & (0.92) & 0.0493 & (0.0018) & 0.0483 & (0.0019) & 52.7\% & (2.98) & 57.0\% & (2.85) \\
Random, uniform   & 89.9\% & (0.24) & 90.1\% & (0.18) & 0.0451 & (0.0005) & 0.0490 & (0.0010) & 76.8\% & (0.64) & 80.1\% & (0.25) \\
Random, clustered & 76.3\% & (0.24) & 83.1\% & (0.54) & 0.0864 & (0.0011) & 0.0690 & (0.0026) & 52.1\% & (0.61) & 66.9\% & (0.90) \\
\bottomrule
\end{tabular}%
\end{sidewaystable}

\begin{sidewaystable}[htbp]
\centering
\caption{PICP, QICE, and worst-10\% site coverage on KAUST 2b-7 ($S = 10{,}000$, basis $[25,81,121]$; mean (SE), 10 replicates): STDKGC vs.\ Ours.}
\label{tab:cov-2b7}
\begin{tabular}{l r@{$\;$}l r@{$\;$}l r@{$\;$}l r@{$\;$}l r@{$\;$}l r@{$\;$}l}
\toprule
Scenario & \multicolumn{2}{c}{PICP (STDKGC)} & \multicolumn{2}{c}{PICP (Ours)} & \multicolumn{2}{c}{QICE (STDKGC)} & \multicolumn{2}{c}{QICE (Ours)} & \multicolumn{2}{c}{W10 (STDKGC)} & \multicolumn{2}{c}{W10 (Ours)} \\
\midrule
Fixed, uniform    & 87.6\% & (0.12) & 88.3\% & (0.06) & 0.0490 & (0.0008) & 0.0500 & (0.0005) & 75.4\% & (0.21) & 73.9\% & (0.21) \\
Fixed, clustered  & 82.3\% & (0.21) & 87.0\% & (0.17) & 0.0641 & (0.0005) & 0.0613 & (0.0010) & 65.3\% & (0.42) & 70.5\% & (0.43) \\
Random, uniform   & 89.9\% & (0.05) & 90.4\% & (0.05) & 0.0429 & (0.0005) & 0.0443 & (0.0005) & 83.4\% & (0.08) & 81.9\% & (0.12) \\
Random, clustered & 84.2\% & (0.14) & 90.5\% & (0.10) & 0.0620 & (0.0011) & 0.0544 & (0.0012) & 73.0\% & (0.28) & 81.3\% & (0.18) \\
\bottomrule
\end{tabular}%
\hskip0.3cm
\centering
\caption{PICP, QICE, and worst-10\% site coverage on KAUST 2b-9 ($S = 10{,}000$, basis $[25,81,121]$; mean (SE), 10 replicates): STDKGC vs.\ Ours.}
\label{tab:cov-2b9}
\begin{tabular}{l r@{$\;$}l r@{$\;$}l r@{$\;$}l r@{$\;$}l r@{$\;$}l r@{$\;$}l}
\toprule
Scenario & \multicolumn{2}{c}{PICP (STDKGC)} & \multicolumn{2}{c}{PICP (Ours)} & \multicolumn{2}{c}{QICE (STDKGC)} & \multicolumn{2}{c}{QICE (Ours)} & \multicolumn{2}{c}{W10 (STDKGC)} & \multicolumn{2}{c}{W10 (Ours)} \\
\midrule
Fixed, uniform    & 98.7\% & (0.03) & 98.7\% & (0.04) & 0.1125 & (0.0005) & 0.1180 & (0.0005) & 94.0\% & (0.19) & 93.7\% & (0.25) \\
Fixed, clustered  & 97.0\% & (0.09) & 97.5\% & (0.05) & 0.0958 & (0.0006) & 0.0982 & (0.0006) & 85.8\% & (0.62) & 88.2\% & (0.30) \\
Random, uniform   & 98.6\% & (0.02) & 98.5\% & (0.02) & 0.1125 & (0.0006) & 0.1169 & (0.0004) & 95.1\% & (0.06) & 94.8\% & (0.06) \\
Random, clustered & 94.7\% & (0.02) & 96.5\% & (0.06) & 0.0854 & (0.0003) & 0.0914 & (0.0005) & 80.4\% & (0.20) & 89.0\% & (0.21) \\
\bottomrule
\end{tabular}%

\end{sidewaystable}

\paragraph{QICE.} 
The QICE and  Worst-10\% site coverage are informed in Table \ref{tab:cov-2a7} - \ref{tab:cov-2b9}.
QICE measures distributional calibration uniformity. 
Ours reduces QICE in most clustered scenarios, indicating better-calibrated quantile intervals:
\begin{itemize}
  \item 2a-8 random-clustered: 0.0790 $\to$ 0.0505.
  \item 2a-9 random-clustered: 0.0864 $\to$ 0.0690.
  \item 2b-7 random-clustered: 0.0620 $\to$ 0.0544.
\end{itemize}
 In uniform scenarios, QICE differences are small, and occasionally Ours shows slightly higher QICE
(e.g., 2b-9 fixed-uniform: 0.1125 $\to$ 0.1180), likely because the global conformal approach
is already well-suited when spatial heterogeneity is low.

\paragraph{Worst-10\% site coverage.}
This metric captures tail reliability---the mean coverage of the 10\% worst-covered sites.
The improvements are most pronounced under clustered observations:
\begin{itemize}
  \item 2a-9 random-clustered: 52.1\% $\to$ 66.9\% (66.9\% - 52.1\% = $+$14.8\%).
  \item 2a-8 random-clustered: 68.1\% $\to$ 79.7\% (79.7\% - 68.1\%= $+$11.6\%).
  \item 2b-7 random-clustered: 73.0\% $\to$ 81.3\% (81.3\% - 73.0\% = $+$8.3\%).
  \item 2b-9 random-clustered: 80.4\% $\to$ 89.0\% (89.0\% - 80.4\% = $+$8.6\%).
  \item 2a-7 random-clustered: 76.9\% $\to$ 81.4\% (81.4\% - 76.9\% = $+$4.5\%).
\end{itemize}
These gains confirm that cluster-aware CQR specifically targets the spatially disadvantaged sites
that global conformal methods fail to protect.

\begin{enumerate}
  \item \textbf{Clustered $>$ uniform benefit}: Across all five datasets, the improvement from
        DA-STDK-MQ  + cluster-aware CQR is largest in clustered scenarios and smallest (sometimes
        negligible or slightly negative) in uniform ones. This validates the design motivation.
  \item \textbf{Process difficulty}: Process 9 datasets (2a-9, 2b-9) exhibit the widest performance
        gap between clustered and uniform settings, suggesting a spatially heterogeneous underlying
        process that particularly benefits from adaptive methods.
  \item \textbf{Grid size effect}: The 2a datasets ($S = 1{,}000$) show lower absolute PICP and W10
        than 2b ($S = 10{,}000$), especially in fixed-site scenarios. With fewer sites, each spatial
        cluster has fewer calibration samples, reducing conformal correction precision.
        Despite this, the relative improvement from Ours remains substantial.
\end{enumerate}
\end{appendices}






\end{document}